\documentclass[letterpaper,journal]{IEEEtran}
\usepackage{url}
\usepackage{xcolor}
\usepackage{cite,graphicx,amsmath,amssymb}
\usepackage{subfigure}
\usepackage{fancyhdr}
\usepackage{mdwmath}
\usepackage{mdwtab}
\usepackage{caption}
\usepackage{amsthm}
\usepackage{setspace}
\usepackage{hyperref}
\usepackage{algorithm}
\usepackage{algorithmic}
\usepackage{multirow}
\usepackage{makecell}
\usepackage{cuted}
\usepackage{mathtools}
\usepackage{mathrsfs}
\usepackage{bm}
\usepackage{pifont}
\hypersetup{colorlinks=false}
\usepackage{stfloats}

\graphicspath{{figs/}}

\newtheorem{theorem}{Theorem}
\newtheorem{proposition}{Proposition}
\newtheorem{lemma}{Lemma}

\allowdisplaybreaks
\DeclareMathOperator*{\argmax}{arg\,max}

\DeclareMathOperator*{\arglexmax}{arg\,lex\,max}
\DeclareMathOperator{\diag}{diag}
\DeclareMathOperator{\rank}{rank}

\newcommand{\Cbb}{\mathbb{C}}

\newcommand{\Zbb}{\mathbb{Z}}
\newcommand{\Acal}{\mathcal{A}}
\newcommand{\Bcal}{\mathcal{B}}
\newcommand{\Ccal}{\mathcal{C}}
\newcommand{\Dcal}{\mathcal{D}}

\newcommand{\Fcal}{\mathcal{F}}
\newcommand{\Gcal}{\mathcal{G}}
\newcommand{\Ical}{\mathcal{I}}
\newcommand{\Kcal}{\mathcal{K}}
\newcommand{\Mcal}{\mathcal{M}}
\newcommand{\Ncal}{\mathcal{N}}
\newcommand{\Pcal}{\mathcal{P}}

\newcommand{\Rcal}{\mathcal{R}}
\newcommand{\Scal}{\mathcal{S}}
\newcommand{\Vcal}{\mathcal{V}}

\newcommand{\bb}{\mathbf{b}}
\newcommand{\cb}{\mathbf{c}}

\newcommand{\gb}{\mathbf{g}}
\newcommand{\hb}{\mathbf{h}}
\newcommand{\pb}{\mathbf{p}}
\newcommand{\Pb}{\mathbf{P}}
\newcommand{\qb}{\mathbf{q}}

\newcommand{\vb}{\mathbf{v}}
\newcommand{\zb}{\mathbf{z}}
\newcommand{\Cb}{\mathbf{C}}
\newcommand{\Db}{\mathbf{D}}
\newcommand{\Gb}{\mathbf{G}}
\newcommand{\Ib}{\mathbf{I}}
\newcommand{\Nb}{\mathbf{N}}
\newcommand{\Rb}{\mathbf{R}}
\newcommand{\Sb}{\mathbf{S}}
\newcommand{\Yb}{\mathbf{Y}}
\newcommand{\Jb}{\mathbf{J}}

\newcommand{\Zb}{\mathbf{Z}}
\newcommand{\ib}{\mathbf{i}}
\newcommand{\tb}{\mathbf{t}}
\newcommand{\SINR}{\mathrm{SINR}}

\newcommand{\neff}{n_{\scriptscriptstyle\mathrm{eff}}}
\newcommand{\CN}{\mathcal{CN}}
\newcommand{\ind}{\mathbb{I}}

\expandafter\def\expandafter\normalsize\expandafter{%
\normalsize%
\setlength\abovedisplayskip{2pt}%
\setlength\belowdisplayskip{2pt}%
\setlength\abovedisplayshortskip{2pt}%
\setlength\belowdisplayshortskip{2pt}%
}

\newcommand{\Require}{\REQUIRE}
\newcommand{\Ensure}{\ENSURE}
\begin{document}

\title{Reconfigurable-Aperture Codebooks and Joint Beam Training for Pinching-Antenna Systems}

\author{
Cong Yu,
Wei Chen,~\IEEEmembership{Senior Member,~IEEE},
Xu Gan,~\IEEEmembership{Member,~IEEE},
Bo Ai,~\IEEEmembership{Fellow,~IEEE},\\
Xiongkuo Min,~\IEEEmembership{Member,~IEEE},
Guangtao Zhai,~\IEEEmembership{Fellow,~IEEE},
Yuanwei Liu,~\IEEEmembership{Fellow,~IEEE}

\thanks{ Cong Yu, Wei Chen and Bo Ai are with School of Electronic
 and Information Engineering, Beijing Jiaotong University, Beijing 100044,
 China. (e-mail: \{cong\_yu, weich, boai\}@bjtu.edu.cn). Xu Gan and Yuanwei Liu are with the Department of Electrical and Electronic Engineering, The University of Hong Kong, Hong Kong. (e-mail: \{eee.ganxu, yuanwei\}@hku.hk). Xiongkuo Min and
Guangtao Zhai are with the Institute of Image Communication and Network
Engineering, Shanghai Jiao Tong University, Shanghai 200240, China. (e-mail: \{minxiongkuo, zhaiguangtao\}@sjtu.edu.cn).}
 }
\maketitle
\pagestyle{empty}
\thispagestyle{empty}

\begin{abstract}
Pinching-antenna systems (PASS) enable flexible aperture reconfiguration by adjusting both the positions and number of active pinching antennas (PAs). Such flexibility gives rise to configuration-dependent spatial coverage patterns. As a result, conventional codebooks designed for fixed-aperture arrays become unsuitable. To address this issue, a unified framework for reconfigurable-aperture codebook design and joint beam training is proposed in this paper. Specifically, a spatial coverage criterion is established to characterize the coverage capability of different PA configurations. By jointly optimizing PA positions and number of active PAs and eliminating redundant configurations, a compact codebook with guaranteed coverage of the entire service region is constructed. Building upon the resulting spatial coverage information, a joint beam training scheme is developed to efficiently select PA configurations across multiple waveguides for multi-user transmission, circumventing exhaustive configuration search. An off-grid refinement strategy is further developed to enhance transmission performance beyond the discrete codebook. Numerical results show that: 1) the proposed compact codebook achieves full spatial coverage with a channel power loss of no more than 3 dB relative to the optimal PA configuration; 2) under the same training and power budgets, it achieves rate gains of 2–5 bit/s/Hz over existing PASS training and conventional near-field codebooks; and 3) the proposed joint beam training scheme exhibits favorable scalability, achieving substantial per-user rate gains as the number of users and waveguides increase.
\end{abstract}


% =====================================================================
\section{Introduction}
\label{sec:intro}
\IEEEPARstart{S}{ixth}-generation (6G) wireless networks are envisioned to provide high-rate and reliable connectivity for densely distributed users over large service areas \cite{intro6G1,intro6G2,intro3}. To meet these demands, reconfigurable antenna architectures have attracted increasing attention by adapting the radiating aperture to the propagation environment \cite{intro1,intro2}. Among them, pinching-antenna systems (PASS) realize flexible aperture reconfiguration through dielectric waveguides and position-adjustable pinching antennas (PAs) \cite{pass1}. Specifically, signals propagate inside each waveguide and are radiated through activated PAs \cite{pass2,pass3}. Through this architecture, both the PA positions and their number can be changed, providing additional flexibility for wireless transmission. Accordingly, selecting a suitable PA configuration becomes an important problem, since the communication gain depends on both the PA configuration and user location \cite{pass5, pass6}. However, the user location or complete channel state information is generally unavailable in advance, while the continuous PA positions and variable number of active PAs make exhaustive configuration search impractical.

Motivated by this challenge, beam training provides a practical approach for identifying a suitable PA configuration through a finite number of measurements. However, conventional codebooks are designed for fixed antenna apertures and typically constructed by discretizing the angle and/or distance domains. Such designs cannot adequately capture the configuration-dependent spatial coverage arising from aperture reconfiguration in PASS. Moreover, in multi-waveguide PASS, the PA configurations of different waveguides need to be jointly determined to support multi-user transmission, resulting in a combinatorial growth of the search space. These challenges motivate a new codebook design that explicitly characterizes the spatial coverage of individual PA configurations and facilitates efficient joint beam training across multiple waveguides.

\subsection{Prior Work}
Motivated by the need to synthesize favorable PA configurations, acquire channels efficiently, and organize a finite training set, existing studies on codebook design and beam training are reviewed along three related lines.

\subsubsection{Codebooks and Beam Training for Fixed-Antenna Arrays}
In near-field communications, spherical wavefronts couple user angle and distance, greatly expanding the beam search space compared with far-field channels. To reduce the pilot overhead of searching this two-dimensional (2D) space, Cui \emph{et al.} \cite{neartrain3} sampled the angle-distance domain non-uniformly by minimizing column coherence in the polar domain for compressed channel estimation. Chen \emph{et al.} \cite{neartrain1} designed multi-resolution hierarchical codebooks to narrow the angle-distance search region layer by layer across the Fresnel region, and Wu \emph{et al.} \cite{neartrain2} used the power pattern of discrete Fourier transform (DFT) beams to jointly estimate user angle and off-grid distance. While these methods reduce the training overhead for fixed arrays, their polar-domain organization assumes a constant physical aperture across all codewords. 

\subsubsection{Codebooks and Beam Training for Movable-Antenna Systems}
Movable-antenna and fluid-antenna systems adjust element positions to exploit local channel variations. To reduce this search cost, Kang \cite{kangMAMulticast} trained deep networks to map pilot observations directly to movable-array positions for multicast transmission, and Shao \emph{et al.} \cite{shao6DMA} developed directional-sparsity maps relating array placement to multipath directions. Moreover, coarse-to-fine port probing was proposed for fluid antennas \cite{pengFASBeamTraining}, multi-position pilot probing was designed for movable holographic arrays \cite{zhangMovableHolographic}, and hierarchical angle-position search was developed for movable subarrays \cite{xiangMABeamTraining}. Although these methods reduce probing overhead for locally movable elements, their candidate sets are restricted to wavelength-scale movement regions with a fixed number of active elements. 

\subsubsection{Beamforming and Beam Training for PASS}
Beamforming and beam training for PASS have mainly been studied from the perspective of configuration optimization, with beamforming designs typically assuming that channel or location information is available. Continuous and discrete activation models were jointly optimized with digital beamforming for PA position design \cite{wangPASSModel}, while hardware-aware designs further optimized the number of active PAs and radiation state \cite{xuPASSPowerModel}. Capacity and outage- and rate-oriented studies further quantified the benefits of PA adaptation under known channel conditions \cite{ouyangPASSCapacity,tyrovolasPASSPerformance}, and a Karush-Kuhn-Tucker-guided Transformer was developed to map a known channel to a PASS configuration \cite{xuPASSKKT}. In comparison, beam training for PASS has received much less attention. Existing studies have considered sequential configuration search, including three-stage beam training (3SBT) for geometry acquisition, phase matching, and PA activation refinement \cite{pass_train}, as well as multimodal learning for ranking PA activations and selecting subsequent probes \cite{ganPASSLLMTraining}. However, these methods do not explicitly exploit the distinctive spatial response characteristics of PASS, thus limiting the efficiency and systematic design of PASS beam training.

\subsection{Motivation and Contributions}
Existing studies on PASS have primarily focused on PA configuration optimization under known channel conditions. The few works addressing beam training rely on sequential configuration search. Unlike fixed-aperture arrays, PASS can reshape its spatial coverage by adjusting the positions and number of active PAs. Consequently, different PA configurations are suitable for serving different spatial regions. Nevertheless, existing beam training methods primarily focus on searching and refining individual PA configurations without explicitly considering their spatial coverage. This not only leads to potentially redundant candidate configurations but also limits the efficiency of configuration selection during beam training. These observations motivate a spatial coverage-based codebook design that reduces the configuration search space and facilitates efficient joint beam training in PASS.

Building upon these observations, we propose a unified framework for reconfigurable-aperture codebook design and joint beam training in PASS. The key idea is to exploit the configuration-dependent spatial coverage of PASS to construct a compact codebook with guaranteed coverage of the entire service region. The resulting spatial coverage information is further leveraged to facilitate efficient joint PA configuration selection across multiple waveguides with reduced training overhead. The main contributions are summarized as follows.
\begin{itemize}
\item \emph{PASS coverage formulation and finite certificate:} For each subregion, a PA configuration is designed to maintain high gain over a target region that encloses it and to limit leakage beyond a guard margin. A finite certificate is then derived to verify coverage throughout the subregion using analytical bounds on the amplitudes and phases of the PA responses.
\item \emph{Reconfigurable-aperture codebook construction:} We design a PA configuration for each subregion by jointly optimizing its PA positions and number of active PAs. To reduce the search dimension, the PA positions are represented by one common phase and a set of discrete indices, while the physical positions remain continuous. The generated configurations are evaluated over all subregions to determine the regions they can serve. Greedy coverage selection then removes redundant configurations while preserving complete coverage.

\item \emph{Codebook-based joint beam training:} We develop a three-stage joint beam-training procedure that exploits the stored spatial information of codewords. It combines location-guided adaptive probing with joint configuration search using measured and stored channel information, followed by PA-group-based off-grid position refinement and final candidate confirmation. This avoids exhaustive probing and testing of all joint configurations.

\item \emph{Performance evaluation:} Numerical results demonstrate that: 1) the proposed compact codebook achieves full spatial coverage with a channel power loss of no more than 3 dB relative to the optimal PA configuration; 2) under the same training and power budgets, it achieves rate gains of 2–5 bit/s/Hz over existing PASS training and conventional near-field codebooks; and 3) the proposed joint beam training scheme exhibits favorable scalability, achieving substantial per-user rate gains as the numbers of users and waveguides increase.
\end{itemize}

\subsection{Organization and Notation}
The remainder of this paper is structured as follows. Section~\ref{sec:model} presents the PASS channel and formulates the coverage requirement. Section~\ref{sec:pa_design} generates one candidate PA configuration for each subregion. Section~\ref{sec:codebook} constructs the shared codebook and develops the joint configuration construction algorithm. Section~\ref{sec:training} applies this algorithm to the measurements and location estimates obtained during beam training. Sections~\ref{sec:results} and \ref{sec:conclusion} report the numerical results and conclusions.

\emph{Notation:} Bold letters denote vectors, matrices, or configuration tuples. Superscripts $(\cdot)^T$, $(\cdot)^H$, and $(\cdot)^*$ denote transpose, Hermitian transpose, and complex conjugation, respectively. The symbols $\|\cdot\|_2$, $\rank(\cdot)$, $\kappa(\cdot)$, and $\lambda_{\min}(\cdot)$ denote the Euclidean or spectral norm, rank, condition number, and minimum eigenvalue. Also, $[x]_+=\max\{x,0\}$ and $\ind\{\cdot\}$ is the indicator function. For a positive integer $K$, $[K]=\{1,\ldots,K\}$ and $\Ib_K$ is the $K\times K$ identity matrix; $[0]=\varnothing$. The operator $\arglexmax$ denotes the set of lexicographic maximizers.

% =====================================================================
\section{System Model and Problem Formulation}
\label{sec:model}
% =====================================================================
This section establishes the signal model and introduces a coverage-driven formulation for PASS codebook design. We first present the multi-waveguide PASS channel model and the performance metric used to evaluate joint configurations. We then show that PASS configurations induce distinct spatial coverage regions, which motivates organizing codewords by the regions they can cover. Finally, we formulate a minimum-cardinality coverage problem that serves as the design objective of the proposed codebook.

\subsection{Channel Model and Performance Metric}

\begin{figure}[t]
\centering
\includegraphics[width=0.8\columnwidth]{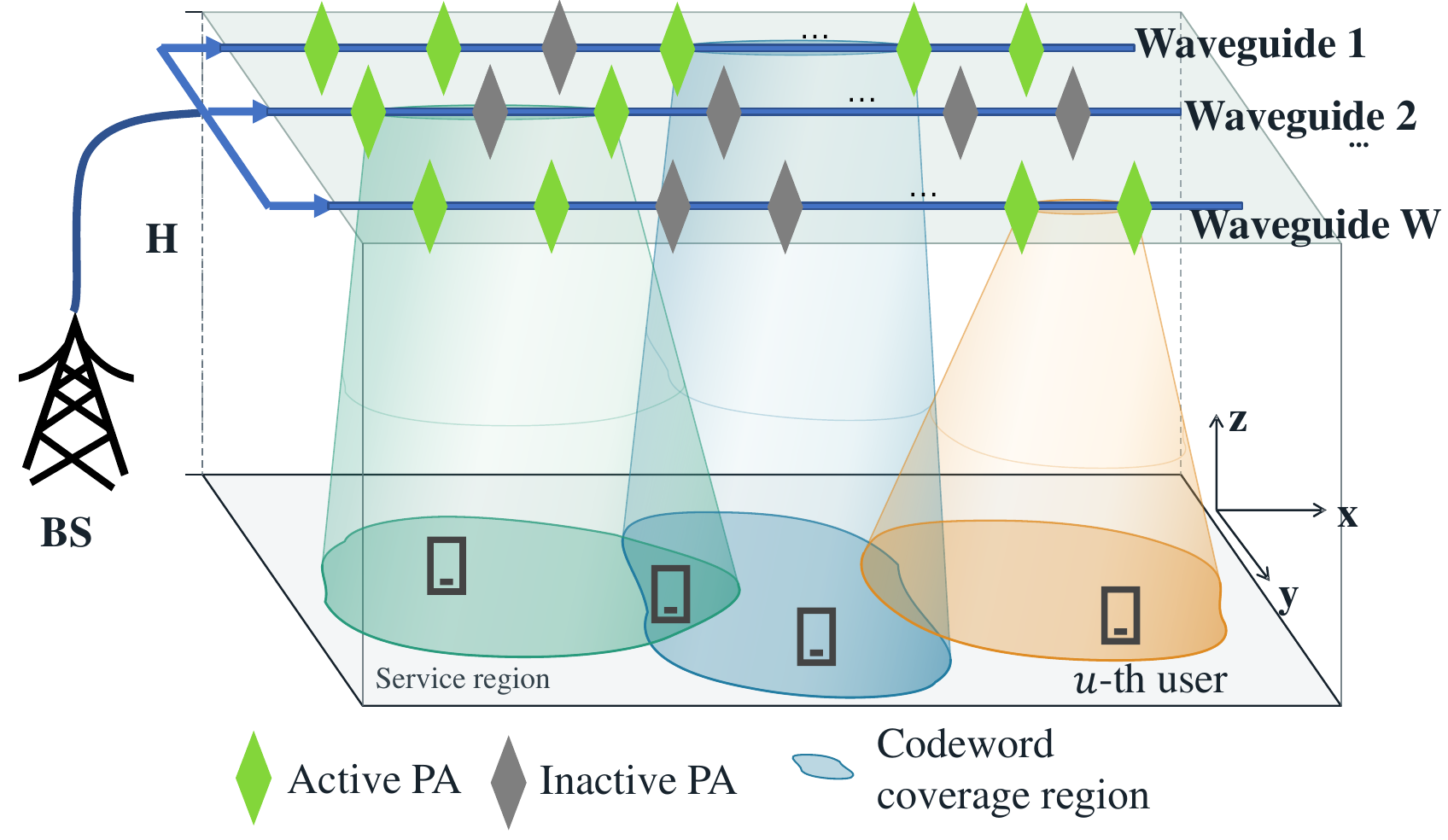}
\caption{Illustration of a multi-waveguide multi-user PASS and the proposed codebook-based joint beam training architecture.}
\label{fig:system}
\end{figure}
As shown in Fig.~\ref{fig:system}, we consider a multi-waveguide multi-user (MWMU) scenario, where $W$ dielectric waveguides of length $L$ are deployed parallel to the $x$-axis at height $H$. Each waveguide is driven by an independent radio-frequency (RF) chain fed at $\xi=0$. A PA activated at longitudinal coordinate $\xi\in[0,L]$ on waveguide $w\in[W]$ has the position $[\xi,y_w,H]^T$, where $y_w$ is the transverse coordinate of waveguide $w$. On the $w$-th waveguide, $M_w$ PAs are active and their ordered longitudinal coordinates are collected in $\boldsymbol\xi_w=(\xi_{w,1},\ldots,\xi_{w,M_w})$. Thus, each $\xi_{w,g}$ is a scalar coordinate and $\zb_w=(M_w,\boldsymbol\xi_w)$ is a PA configuration. Subject to a maximum of $N$ active PAs and a minimum inter-PA spacing $d_{\min}>0$ to avoid mutual coupling, the set of feasible single-waveguide PA configurations is
\begin{equation}
\Fcal_N=
\left\{(M_w,\boldsymbol\xi_w)\;\middle|\;
\begin{array}{@{}l@{}}
M_w\in[N],\\
\boldsymbol\xi_w\in[0,L]^{M_w},\\
\xi_{w,g+1}\!-\!\xi_{w,g}\!\ge d_{\min}, g\in[M_w\!-\!1].
\end{array}\right\}
\label{eq:feasible}
\end{equation}
Across all $W$ waveguides, the joint configuration is the ordered tuple $\Zb=(\zb_1,\ldots,\zb_W)\in\Fcal_N^W$.

The system simultaneously serves $U\le W$ single-antenna users in the ground service region $\Omega_{xy}=[x_{\min},x_{\max}]\times[y_{\min},y_{\max}]$, where user $u\in[U]$ is located at $\pb_u=[x_u,y_u,0]^T$. Since waveguide $w$ runs along the $x$-axis at $(y_w,H)$, the distance from any PA on waveguide $w$ to a ground point $(x,y)$ depends on $y$ only through the transverse distance $\rho_w(y)=\sqrt{(y-y_w)^2+H^2}$. We therefore represent any ground location $(x,y)$ relative to waveguide $w$ by the 2D waveguide-relative coordinate
\begin{equation}
\qb_w(x,y)=\left[x,\sqrt{(y-y_w)^2+H^2}\right]^T,
\label{eq:reduced_coordinate_map}
\end{equation}
with $\qb_w(\pb_u)=[x_u,\rho_w(y_u)]^T$ for user $u$, and write $\qb=[x,\rho]^T$ for a generic waveguide-relative coordinate. Expressing channels in $(x,\rho)$ places all structurally identical waveguides under the same single-waveguide spatial response model, allowing us to design a single shared codebook offline and reuse it across all $W$ waveguides.

The PASS channel combines guided propagation inside the waveguide with spherical free-space propagation from each active PA. For a PA at $\xi$, its free-space distance to a point with waveguide-relative coordinates $\qb$ is $r(\xi;\qb)=\sqrt{(\xi-x)^2+\rho^2}$. For the $g$-th active PA on waveguide $w$ and the $u$-th user, the distance is $r(\xi_{w,g};\qb_w(\pb_u))$. Let $\lambda$ denote the free-space wavelength, $k=2\pi/\lambda$ the wavenumber, and $\neff>1$ the waveguide effective refractive index. The guided and free-space distances together determine the phase and the single-PA channel coefficient:
\begin{align}
\phi(\xi;\qb)&=k\big(r(\xi;\qb)+\neff\xi\big),
\label{eq:single_pa_phase}\\
a(\xi;\qb)&=\frac{\eta}{r(\xi;\qb)}\exp[-\mathrm j\phi(\xi;\qb)],
\label{eq:single_pa_response}
\end{align}
where $\eta$ is the reference path-loss factor and $\mathrm j$ is the imaginary unit. With waveguide power split equally among the $M_w$ active PAs, the channel from waveguide $w$ to user $u$ is
\begin{equation}
g_{u,w}(\zb_w;\pb_u)=\frac{1}{\sqrt{M_w}}
\sum_{g=1}^{M_w}a(\xi_{w,g};\qb_w(\pb_u)).
\label{eq:scalar_channel}
\end{equation}
Stacking $g_{u,w}(\zb_w;\pb_u)$ across all $U$ users and $W$ waveguides gives the $U\times W$ effective multi-user channel matrix
\begin{equation}
\Gb(\Zb;\Pb)=
\begin{bmatrix}
 g_{1,1}(\zb_1;\pb_1)&\cdots&g_{1,W}(\zb_W;\pb_1)\\
\vdots&\ddots&\vdots\\
 g_{U,1}(\zb_1;\pb_U)&\cdots&g_{U,W}(\zb_W;\pb_U)
\end{bmatrix},
\label{eq:effective_channel}
\end{equation}
where $\Pb=(\pb_1,\ldots,\pb_U)$ collects the user positions.

For multi-user transmission, we consider linear precoding based on the effective channel $\Gb(\Zb;\Pb)$ and evaluate a joint configuration through its achievable rates. Specifically, the total transmit power is split equally across the $U$ users and zero-forcing (ZF) precoding is applied when the channel estimate $\widehat{\Gb}$ has full row rank. With $\Gb=\Gb(\Zb;\Pb)$ and $(\cdot)^\dagger$ denoting the Moore--Penrose pseudoinverse, the signal-to-interference-plus-noise ratio (SINR) of the $u$-th user is
\begin{equation}
\SINR_u(\Zb;\!\Pb,\!\widehat{\Gb})\!=\!\!
\begin{cases}
\displaystyle
\frac{
|[\Gb\widehat{\Gb}^{\dagger}]_{u,u}|^2/
\|[\widehat{\Gb}^{\dagger}]_{:,u}\|_2^2
}{
\displaystyle\sum_{\ell\ne u}
\frac{|[\Gb\widehat{\Gb}^{\dagger}]_{u,\ell}|^2}
{\|[\widehat{\Gb}^{\dagger}]_{:,\ell}\|_2^2}
\!+\!\frac{U}{\gamma_{\mathrm{tx}}}
},\!\!\!\!\!
& \rank(\widehat{\Gb})\!=\!U,\\
0, & \rank(\widehat{\Gb})\!<\!\!U.
\end{cases}
\label{eq:sinr}
\end{equation}
where $\gamma_{\mathrm{tx}}=P_{\scriptscriptstyle\mathrm{tot}}/\sigma^2$ is the transmit signal-to-noise ratio (SNR), $P_{\scriptscriptstyle\mathrm{tot}}$ is the total transmit power, and $\sigma^2$ is the receiver noise variance. For the ideal reference, the true channel is used as the estimate, i.e., $\widehat{\Gb}=\Gb$. In this case, \eqref{eq:sinr} reduces to $\gamma_{\mathrm{tx}}\|\Gb\|_F^2$ for $U=1$. A joint PA configuration that achieves a high average rate may still leave one user with a low rate. To account for this, we evaluate each configuration using both the average rate and the lowest user rate. With $R_u(\Zb;\Pb)=\log_2\big(1+\SINR_u(\Zb;\Pb,\Gb)\big)$, the reference utility is defined as
\begin{equation}
\mathcal U_{\mathrm{tr}}(\Zb;\Pb)=
\frac1U\sum_{u=1}^{U}R_u(\Zb;\Pb)
+\omega_{\mathrm{tr}}\min_{u\in[U]}R_u(\Zb;\Pb),
\label{eq:true_joint_utility}
\end{equation}
with fairness weight $\omega_{\mathrm{tr}}>0$. Under known user locations, the ideal reference configuration is
\begin{equation}
\Zb^{\mathrm{tr}}(\Pb)\in
\argmax_{\Zb\in\Fcal_N^W}
\mathcal U_{\mathrm{tr}}(\Zb;\Pb).
\label{eq:true_joint_reference}
\end{equation}
However, user locations $\Pb$ are unknown before transmission, and searching the mixed discrete-continuous space $\Fcal_N^W$ directly online is intractable. We therefore divide the overall design into an offline stage and an online stage. Offline, we construct a compact shared single-waveguide codebook $\Ccal\subseteq\Fcal_N$ that guarantees spatial power coverage over the service region. Online, we use limited pilot measurements to select a joint multi-waveguide configuration from the on-grid search space $\Ccal^W$, and optionally perform off-grid continuous PA refinement outside $\Ccal^W$ when the SNR and pilot budget permit.

\subsection{Irregular and Non-Uniform Spatial Coverage Regions}

For a single waveguide, the channel power gain of a PA configuration $\zb_w=(M_w,\boldsymbol\xi_w)$ at waveguide-relative coordinate $\qb$ is
\begin{equation}
\Pcal(\zb_w;\qb)
=\frac{1}{M_w}\left|\sum_{g=1}^{M_w}a(\xi_{w,g};\qb)\right|^2.
\label{eq:single_link_boundary}
\end{equation}
Expanding \eqref{eq:single_link_boundary} via \eqref{eq:single_pa_response} yields
\begin{equation}
\Pcal(\zb_w;\qb)=\frac{1}{M_w}\left[
\sum_{g=1}^{M_w}\frac{\eta^2}{r_g^2}
+2\sum_{1\le g<h\le M_w}\frac{\eta^2\cos(\phi_g-\phi_h)}{r_gr_h}
\right],
\label{eq:pairwise_response}
\end{equation}
where $r_g=r(\xi_{w,g};\qb)$ and $\phi_g=\phi(\xi_{w,g};\qb)$. We refer to the spatial region over which $\Pcal(\zb_w;\qb)$ achieves a prescribed gain threshold as the \emph{coverage region} of $\zb_w$. Unlike fixed-aperture arrays whose beam patterns are governed solely by phase shifts over a constant geometry, reconfiguring $\zb_w$ in \eqref{eq:pairwise_response} simultaneously alters the link distances $\{r_g\}$, the spatial variation of the pairwise phase differences $\{\phi_g-\phi_h\}$, and the $1/M_w$ power-splitting factor along with the number of coherent-sum terms. As a result, the spatial response of PASS is highly non-uniform, and its coverage region becomes highly irregular across configurations.

\begin{figure}[t]
\centering
\includegraphics[width=0.9\columnwidth]{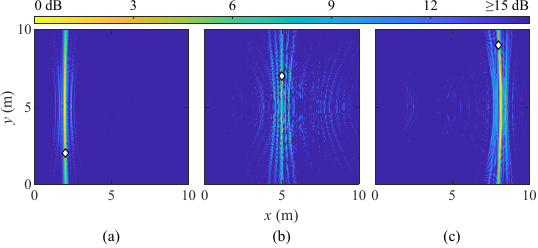}
\caption{Spatial power responses for three configurations with (a) 26, (b) 28, and (c) 24 active PAs. The color scale shows power loss relative to each configuration's peak response, from yellow (0~dB) to dark purple ($\geq 15$~dB). White diamonds mark the target locations.}
\label{fig:pass_response_maps}
\end{figure}

This behavior is illustrated in Fig.~\ref{fig:pass_response_maps} for three representative PA configurations synthesized for distinct target locations. Although each response is mirror-symmetric about the waveguide axis $y=y_w$, the high-gain regions differ markedly in size and shape. In \eqref{eq:pairwise_response}, the path-loss weights $1/(r_gr_h)$ increase as the transverse distance to the waveguide decreases at a fixed $x$. The coherence terms $\cos(\phi_g-\phi_h)$ vary with location and determine where the PA signals reinforce or cancel one another. Together, these terms explain why the high-gain region changes with the PA configuration.

Such highly irregular and spatially non-uniform coverage regions render conventional fixed-grid codebooks (e.g., uniform angle-distance or Cartesian partitions designed for fixed arrays) unsuitable for PASS. Specifically, organizing PASS codewords on a predetermined fixed spatial grid inevitably causes two severe inefficiencies: in some regions, the actual coverage region of an assigned configuration is overly small, leaving coverage gaps or severe gain degradation; in other regions, configurations produce broad coverage regions with heavy mutual overlap, resulting in redundant codewords and wasted beam-training overhead. Therefore, rather than relying on a rigid a priori spatial partition, we consider a coverage-oriented codebook design that explicitly characterizes the actual coverage region of each candidate configuration.

\subsection{Coverage-Oriented Codebook Formulation}

To eliminate the coverage gaps and redundant overlap identified above, our goal is to construct a compact shared single-waveguide codebook $\Ccal=\{\zb_i\}_{i=1}^{Q}\subseteq\Fcal_N$, providing a finite basis for online joint beam training toward $\Zb^{\mathrm{tr}}(\Pb)$ in \eqref{eq:true_joint_reference}. Because all waveguides share the same response model over $(x,\rho)$, $\Ccal$ is designed over the union $\Omega_{\mathrm r}=\bigcup_{w=1}^{W}\Omega_w$, where $\Omega_w=\{\qb_w(x,y):(x,y)\in\Omega_{xy}\}$. Omitting $w$ for single-waveguide design, let $\Pcal^\star(\qb)=\max_{\zb\in\Fcal_N}\Pcal(\zb;\qb)$ denote the maximum achievable gain at $\qb\in\Omega_{\mathrm r}$. Accordingly, we prevent coverage gaps by requiring every $\qb\in\Omega_{\mathrm r}$ to stay within $\delta$~dB ($\tau_\delta=10^{-\delta/10}$) of $\Pcal^\star(\qb)$, while suppressing redundant overlap and training overhead by minimizing the codebook cardinality $|\Ccal|$:
\begin{subequations}
\label{eq:continuous_codebook_problem}
\begin{align}
(\mathsf P_\delta^{\mathrm{CB}})\quad
&\min_{\Ccal\subseteq\Fcal_N}
|\Ccal|\\
\mathrm{s.t.}\quad&
\max_{\zb\in\Ccal}\Pcal(\zb;\qb)
\ge\tau_\delta \Pcal^\star(\qb),
\quad \forall\qb\in\Omega_{\mathrm r}.
\label{eq:complete_coverage}
\end{align}
\end{subequations}
Solving $(\mathsf P_\delta^{\mathrm{CB}})$ directly is intractable because $\Fcal_N$ is a mixed discrete-continuous space and \eqref{eq:complete_coverage} must hold at infinitely many continuous points $\qb\in\Omega_{\mathrm r}$. To address these challenges, Sections~\ref{sec:pa_design} and~\ref{sec:codebook} decompose the problem into two stages. In Section~\ref{sec:pa_design}, the waveguide-relative design domain is partitioned into smaller subregions, and a high-gain PA configuration is designed for each subregion by jointly optimizing the PA positions and number of active PAs through a phase-aligned parameterization. In Section~\ref{sec:codebook}, an analytical coverage condition is established to determine whether a candidate configuration can serve an entire subregion within the prescribed gain loss.
% =====================================================================
\section{PA Configuration Design}
\label{sec:pa_design}
% =====================================================================

This section develops a structured synthesis method that generates a high-quality candidate pool for Problem~$(\mathsf P_\delta^{\mathrm{CB}})$. We partition the waveguide-relative domain $\Omega_{\mathrm r}$ into $V$ subregions and design one representative configuration for each subregion's center. The objective is to maintain a sufficiently high channel gain throughout a target region enclosing the subregion while limiting the radiation toward the surrounding area. To achieve this, we first establish a region-based optimization criterion and then exploit phase alignment at the design center, and jointly optimize the PA positions and number of active PAs through local search.

\subsection{Subregion-Based PA Design Criterion}

Partition $\Omega_{\mathrm r}$ into $V$ subregions $\{\Vcal_n\}_{n=1}^{V}$, where subregion $\Vcal_n$ has center $\qb_n$ and radius $r_n^{\mathrm{sr}}=\sup_{\qb\in\Vcal_n}\|\qb-\qb_n\|_2$. For subregion $\Vcal_n$, we set the design center to $\cb=\qb_n=[c_x,c_\rho]^T$. Maximizing channel gain solely at $\cb$ often leaves sharp dips near the subregion boundary; a well-shaped configuration should instead sustain high gain across the entire subregion while suppressing excess gain outside it. Thus, we seek high gain across a target region enclosing the subregion and limit leakage beyond a guard margin. First, we define a target region $\Rcal_{\cb}$ enclosing $\Vcal_n$ ($r_{\mathrm R}\ge r_n^{\mathrm{sr}}$) and an out-of-region set $\Scal_{\cb}$ separated by a guard margin $r_{\mathrm G}>0$ to allow smooth spatial roll-off:
\begin{align}
\Rcal_{\cb}&=\Omega_{\mathrm r}\cap
\{\qb:\|\qb-\cb\|_2\le r_{\mathrm R}\},
\label{eq:target_region}\\
\Scal_{\cb}&=\Omega_{\mathrm r}\setminus
\{\qb:\|\qb-\cb\|_2<r_{\mathrm R}+r_{\mathrm G}\}.
\label{eq:out_of_region_set}
\end{align}
Choosing $r_{\mathrm R}>r_n^{\mathrm{sr}}$ further encourages adjacent candidates to form overlapping coverage.

To shape the spatial response over $(\Rcal_{\cb},\Scal_{\cb})$, we evaluate each configuration $\zb\in\Fcal_N$ via the four-component vector
\begin{equation}
\Jb_{\cb}(\zb)\!=\!\left[
\underline{\Pcal}_{\cb}(\zb),\!
-V^{\scriptscriptstyle\mathrm{out}}_{\cb}(\zb),\!
\bar{\Pcal}_{\cb}(\zb),
-\Pcal^{\scriptscriptstyle\mathrm{out}}_{\cb}(\zb)
\right]^{T}\!\!\!,
\label{eq:regional_comparison_vector}
\end{equation}
where, with normalized area $\mu(\Acal)=|\Acal|/|\Omega_{\mathrm r}|$,
\begin{align}
\underline{\Pcal}_{\cb}(\zb)
&=\min_{\qb\in\Rcal_{\cb}}\Pcal(\zb;\qb),
\label{eq:min_in_region_gain}\\
V^{\scriptscriptstyle\mathrm{out}}_{\cb}(\zb)
&=\frac{1}{\mu(\Scal_{\cb})}\int_{\Scal_{\cb}}
\left[\Pcal(\zb;\qb)-\underline{\Pcal}_{\cb}(\zb)\right]_+d\mu(\qb),
\label{eq:average_out_region_excess}\\
\bar{\Pcal}_{\cb}(\zb)
&=\frac{1}{\mu(\Rcal_{\cb})}\int_{\Rcal_{\cb}}
\Pcal(\zb;\qb)d\mu(\qb),
\label{eq:mean_in_region_gain}\\
\Pcal^{\scriptscriptstyle\mathrm{out}}_{\cb}(\zb)
&=\max_{\qb\in\Scal_{\cb}}\Pcal(\zb;\qb).
\label{eq:max_out_region_gain}
\end{align}
Compared lexicographically, $\Jb_{\cb}(\zb)$ first improves the weakest gain $\underline{\Pcal}_{\cb}(\zb)$ in $\Rcal_{\cb}$, next limits the average out-of-region excess $V^{\scriptscriptstyle\mathrm{out}}_{\cb}(\zb)$ above $\underline{\Pcal}_{\cb}(\zb)$, then raises the mean in-region gain $\bar{\Pcal}_{\cb}(\zb)$, and finally suppresses the peak sidelobe $\Pcal^{\scriptscriptstyle\mathrm{out}}_{\cb}(\zb)$. The local PA design problem is formulated as
\begin{equation}
\zb^\star(\cb)\in\arglexmax_{\zb\in\Fcal_N}\Jb_{\cb}(\zb).
\label{eq:local_pa_problem}
\end{equation}

\subsection{Phase-Aligned Parameterization of PA Positions}

Directly solving \eqref{eq:local_pa_problem} over $(M,\boldsymbol\xi)\in\Fcal_N$ is challenging due to rapid carrier-phase variations and the variable dimension $M$. To reduce search dimensionality while retaining continuous PA placement, we restrict attention to configurations whose active PAs combine coherently at $\cb$, aligning each candidate location $\xi_b(a_0)\in[0,L]$ to a common reference phase $a_0\in[0,2\pi)$ modulo $2\pi$:
\begin{equation}
\phi(\xi_b(a_0);\cb)=a_0+2\pi b, \qquad b\in\Zbb,
\label{eq:phase_branch}
\end{equation}
where $b$ indexes phase cycles along the waveguide. Since $\partial\phi(\xi;\cb)/\partial\xi=k\big(\neff+(\xi-c_x)/r(\xi;\cb)\big)>k(\neff-1)>0$ for $\neff>1$, $\phi(\xi;\cb)$ increases strictly monotonically on $[0,L]$. Fixing $a_0$ therefore determines at most one continuous coordinate per cycle $b$, while varying $a_0$ shifts all candidate positions smoothly along the waveguide.

\begin{proposition}[Closed-Form Phase-Aligned PA Locations]
\label{prop:root_map}
\emph{Let $S_{b,a_0}=(b+a_0/(2\pi))\lambda$ and $\neff>1$. A coordinate $\xi_b(a_0)\in[0,L]$ satisfies \eqref{eq:phase_branch} if and only if $b\in[b_{\min},b_{\max}]$, where
\begin{align}
b_{\min}&=\left\lceil
\frac{\sqrt{c_x^2+c_\rho^2}}{\lambda}-\frac{a_0}{2\pi}
\right\rceil,
\label{eq:root_label_min}\\
b_{\max}&=\left\lfloor
\frac{\neff L+\sqrt{(L-c_x)^2+c_\rho^2}}{\lambda}
-\frac{a_0}{2\pi}\right\rfloor,
\label{eq:root_label_max}
\end{align}
and is uniquely given in closed form by
\begin{equation}
\xi_b(a_0)\!=\!\frac{\neff S_{b,a_0}\!-\!c_x\!
-\!\!\sqrt{(S_{b,a_0}-\neff c_x)^2\!+\!(\neff^2-1)c_\rho^2}}{\neff^2\!-\!1}.
\label{eq:root_quadratic}
\end{equation}
Moreover, $\xi_b(a_0)$ varies continuously with $a_0$ on the reference-phase interval over which $\xi_b(a_0)\in[0,L]$.}
\end{proposition}

\begin{proof}
Evaluating $\phi(\xi;\cb)$ at $\xi\in\{0,L\}$ together with its strict monotonicity on $[0,L]$ establishes a unique root in $[0,L]$ if and only if $b\in[b_{\min},b_{\max}]$. Substituting \eqref{eq:single_pa_phase} into \eqref{eq:phase_branch} and setting $\xi=\xi_b(a_0)$ gives $\sqrt{(\xi-c_x)^2+c_\rho^2}=S_{b,a_0}-\neff\xi\ge0$, and squaring yields $(\neff^2-1)\xi^2+2(c_x-\neff S_{b,a_0})\xi+S_{b,a_0}^2-c_x^2-c_\rho^2=0$, with roots $\xi_{\pm}=\big(\neff S_{b,a_0}-c_x\pm\sqrt{\Delta_{b,a_0}}\big)/(\neff^2-1)$ and $\Delta_{b,a_0}=(S_{b,a_0}-\neff c_x)^2+(\neff^2-1)c_\rho^2$. Because $c_\rho\ge H>0$ and $\neff>1$, we have $\neff\sqrt{\Delta_{b,a_0}}>|S_{b,a_0}-\neff c_x|$, so $S_{b,a_0}-\neff\xi_+<0$ and $S_{b,a_0}-\neff\xi_->0$. Hence $\xi_-$ is the unique feasible root in \eqref{eq:root_quadratic}, and its continuity in $a_0$ follows directly from \eqref{eq:root_quadratic}.
\end{proof}

\subsection{Joint Optimization of PA Activation and Positions}

Activating every phase-aligned location from Proposition~\ref{prop:root_map} is generally suboptimal. Under equal power splitting in \eqref{eq:scalar_channel}, adding an $(M+1)$-th PA changes the gain at $\qb$ by
\begin{equation}
\Pcal_{M+1}\!-\!\Pcal_M=
\frac{2\Re\{S_M^*a_{M+1}\}+|a_{M+1}|^2}{M+1}
-\frac{|S_M|^2}{M(M+1)},
\label{eq:cardinality_difference}
\end{equation}
where $a_m\!=\!a(\xi_m;\qb)$, $S_M\!=\!\sum_{m=1}^{M}\!a_m$, and $\Pcal_M\!=\!|S_M|^2\!/\!M$. Distant PAs can therefore reduce gain through $1/(M+1)$ power dilution or unfavorable combining across $\Rcal_{\cb}$.

Consequently, we jointly optimize the reference phase and selected phase-cycle index set $\mathbf{s}=(a_0,\Bcal)$:
\begin{equation}
\mathbf{s}^\star\in\arglexmax_{\mathbf{s}\in\mathcal X_{\cb}}
\Jb_{\cb}\big(\zb(\cb;\mathbf{s})\big),
\label{eq:regional_problem}
\end{equation}
where $\zb(\cb;\mathbf{s})=\big(|\Bcal|,\operatorname{sort}\{\xi_b(a_0):b\in\Bcal\}\big)$ and
\begin{equation}
\mathcal X_{\cb}\!\!=\!\!\Big\{\!\mathbf{s}\!=\!\!(a_0,\Bcal)\!\!:\!a_0\!\!\in\![0,\!2\pi),
\Bcal\!\subseteq\!\!\Bcal_{\cb}(a_0),\!\zb(\cb;\mathbf{s})\!\in\!\!\Fcal_N\!\Big\},
\label{eq:phase_support_set}
\end{equation}
with $\Bcal_{\cb}(a_0)=\{b\in\Zbb:b_{\min}\le b\le b_{\max}\}$.

Problem \eqref{eq:regional_problem} is solved via multi-start local search over initial values of $a_0$. At iteration $t$, $\mathbf{s}^{(t)}$ is updated over the union neighborhood $\Ncal(\mathbf{s}^{(t)})$:
\begin{align}
\mathbf{s}^{(t+1)}&\in
\arglexmax_{\mathbf{s}\in\{\mathbf{s}^{(t)}\}\cup\Ncal(\mathbf{s}^{(t)})}
\Jb_{\cb}\big(\zb(\cb;\mathbf{s})\big),
\label{eq:regional_update}\\
\Ncal(\mathbf{s})&=\Ncal_{\scriptscriptstyle\mathrm{set}}(\mathbf{s})\cup\Ncal_a(\mathbf{s}),
\label{eq:joint_neighborhoods}
\end{align}
where $\Ncal_{\scriptscriptstyle\mathrm{set}}(\mathbf{s})$ adds, removes, or swaps one index in $\Bcal$ at fixed $a_0$, and $\Ncal_a(\mathbf{s})$ perturbs $a_0$ within step size $\Delta_a>0$. For each start, $\Bcal$ is initialized from valid roots near $c_x$ subject to $d_{\min}$, and $\Jb_{\cb}$ is evaluated on discrete samples of $(\Rcal_{\cb},\Scal_{\cb})$. In lexicographic comparison, entries of $\Jb_{\cb}$ differing by at most $\epsilon_J$ are treated as ties; whenever no feasible neighbor improves $\mathbf{s}$ beyond $\epsilon_J$, $\Delta_a$ is reduced toward $\Delta_{a,\min}$ until either $\Delta_{a,\min}$ or an iteration limit is reached. Comparing the feasible configurations returned across all starts under the same criterion yields the candidate $\zb^\star$ for $\Vcal_n$, as summarized in Algorithm~\ref{alg:pa_synthesis}. Running Algorithm~\ref{alg:pa_synthesis} over $n\in[V]$ produces the candidate pool $\Dcal_0=\{\zb_j\}_{j=1}^{Q_0}$, where $Q_0=|\Dcal_0|$.

\begin{algorithm}[t]
\caption{PA Configuration Search for a Given Design Center}
\label{alg:pa_synthesis}
\begin{algorithmic}[1]
\Require Center $\cb$, regions $(\Rcal_{\cb},\Scal_{\cb})$, $N$, $d_{\min}$, initial phases, phase steps $(\Delta_{a,0},\Delta_{a,\min})$, iteration limit, and tolerance $\epsilon_J$.
\Ensure PA configuration $\zb^\star$ and its response quantities.
\FOR{each initial phase $a_0$}
\STATE Compute $[b_{\min},b_{\max}]$ via \eqref{eq:root_label_min}-\eqref{eq:root_label_max} and roots $\{\xi_b(a_0)\}$ via \eqref{eq:root_quadratic}.
\STATE Initialize $\Bcal$ from valid roots near $c_x$ satisfying $d_{\min}$; set $\mathbf{s}=(a_0,\Bcal)$ and $\Delta_a=\Delta_{a,0}$.
\WHILE{iteration limit is not reached}
\STATE Search $\Ncal(\mathbf{s})$ in \eqref{eq:joint_neighborhoods} via \eqref{eq:regional_update} under $\epsilon_J$-tie comparison.
\IF{a feasible neighbor improves $\mathbf{s}$ beyond $\epsilon_J$}
\STATE Update $\mathbf{s}$ to the best improving neighbor.
\ELSIF{$\Delta_a>\Delta_{a,\min}$}
\STATE Reduce $\Delta_a$ toward $\Delta_{a,\min}$.
\ELSE
\STATE Terminate the inner search.
\ENDIF
\ENDWHILE
\ENDFOR
\STATE Select $\zb^\star$ as the best feasible configuration across initializations under the same $\epsilon_J$-tie criterion.
\RETURN $\zb^\star$ and its response quantities.
\end{algorithmic}
\end{algorithm}

% =====================================================================
\section{Codebook and Multi-Waveguide Configuration Construction}
\label{sec:codebook}
% =====================================================================

Section~\ref{sec:pa_design} produces one candidate configuration per subregion, which yields a large pool with substantial spatial overlap. Probing such a redundant pool is inefficient, and naively searching all multi-waveguide combinations is infeasible. This section addresses both issues. First, we certify coverage throughout each subregion by bounding amplitude and phase variations. Using this certificate, we prune the candidate pool via greedy weighted set cover and refine the resulting design centers to obtain a compact shared single-waveguide codebook $\Ccal$ with guaranteed coverage of $\Omega_{\mathrm r}$. Second, leveraging the stored codebook responses and coverage sets, we develop a scalable procedure to construct a small joint candidate set $\Ccal_W\subseteq\Ccal^W$.

\subsection{Analytical Coverage Guarantee over Subregions}

To determine which subregions each candidate $\zb_j\in\Dcal_0$ covers, we bound its gain throughout $\Vcal_n$ and compare it with the threshold in \eqref{eq:complete_coverage}, which is relative to the optimal feasible gain $\Pcal^\star(\qb)$. Because each PA's path-loss amplitude and relative phase vary across $\Vcal_n$, we bound the $g$-th PA's normalized amplitude from below by $\beta_{j,g,n}$ and its relative phase deviation from above by $\Theta_{j,g,n}$, as derived in Appendix~\ref{app:cell_coverage}. Combining these bounds yields the analytical normalized gain lower bound
\begin{equation}
\Gamma_{j,n}=
\begin{cases}
\displaystyle\frac{\left(\sum_{g=1}^{M_j}\beta_{j,g,n}\cos\Theta_{j,g,n}\right)^2}{M_jN},
&\displaystyle\max_{g\in[M_j]}\Theta_{j,g,n}\le\frac{\pi}{2},\\
0,&\text{otherwise},
\end{cases}
\label{eq:certificate_power}
\end{equation}
where $\max_{g\in[M_j]}\Theta_{j,g,n}\le\pi/2$ ensures nonnegative coherent projection onto the reference phase. Since every PA satisfies $r(\xi;\qb)\ge\rho$, any feasible configuration with at most $N$ PAs obeys $\Pcal^\star(\qb)\le N\eta^2/\rho^2$. Checking the scalar condition $\Gamma_{j,n}\ge\tau_\delta$ therefore certifies continuous $\delta$-dB coverage relative to $\Pcal^\star(\qb)$ throughout $\Vcal_n$. As a benchmark over the candidate pool, we also define $\Pcal_0(\qb)=\max_{\zb\in\Dcal_0}\Pcal(\zb;\qb)$. Because $\Pcal_0(\qb)\le\Pcal^\star(\qb)$, the same certificate automatically holds relative to $\Pcal_0(\qb)$.

\begin{lemma}[Subregion Coverage Guarantee]
\label{lem:cell_coverage}
\emph{For any feasible configuration $\zb_j\in\Fcal_N$, its channel gain satisfies
\begin{equation}
\Pcal(\zb_j;\qb)\ge\Gamma_{j,n}\frac{N\eta^2}{\rho^2},
\qquad \forall\qb\in\Vcal_n.
\label{eq:certified_power_lower}
\end{equation}
Consequently, if $\Gamma_{j,n}\ge\tau_\delta$, then $\zb_j$ covers the entire subregion $\Vcal_n$ within $\delta$~dB of the optimal feasible gain:
\begin{equation}
\Pcal(\zb_j;\qb)\ge\tau_\delta \Pcal^\star(\qb)
\ge\tau_\delta \Pcal_0(\qb),
\qquad \forall\qb\in\Vcal_n.
\label{eq:cell_coverage}
\end{equation}}
\end{lemma}
The proof is given in Appendix~\ref{app:cell_coverage}.

\subsection{Greedy Codebook Selection and Design-Center Refinement}

Lemma~\ref{lem:cell_coverage} associates each candidate $\zb_j\in\Dcal_0$ with the indices of the subregions it is certified to cover $\Ical_j=\{n\in[V]:\Gamma_{j,n}\ge\tau_\delta\}$. This converts continuous coverage into a discrete set-cover problem that minimizes the number of retained candidates:
\begin{subequations}
\label{eq:set_cover}
\begin{align}
(\mathsf P_{\mathrm{CB}})\quad
&\min_{\{s_j\}_{j=1}^{Q_0}}\ \sum_{j=1}^{Q_0}s_j\\
\mathrm{s.t.}\quad&
\sum_{j:\,n\in\Ical_j}s_j\ge1,
\quad \forall n\in[V],\\
&s_j\in\{0,1\},\quad \forall j\in[Q_0],
\end{align}
\end{subequations}
where $s_j\in\{0,1\}$ indicates whether $\zb_j$ is selected. Let $w_n=|\Vcal_n|/|\Omega_{\mathrm r}|>0$ denote the normalized area weight of subregion $\Vcal_n$, which sums to one over $n\in[V]$. The area fraction certified by $\Ccal\subseteq\Dcal_0$ is $f(\Ccal)=\sum_{n=1}^{V}w_n\,\ind\{\exists\zb_j\in\Ccal:n\in\Ical_j\}$. Under the premise that the candidate pool certifies complete coverage with $f(\Dcal_0)=1$, we avoid the combinatorial complexity of \eqref{eq:set_cover} through greedy selection. Starting from $\Ccal^{(0)}=\varnothing$, step $t$ adds the candidate that maximizes the newly covered area weight:
\begin{align}
j_t &\in \argmax_{j:\zb_j\notin\Ccal^{(t-1)}}
\left[f(\Ccal^{(t-1)}\cup\{\zb_j\})-f(\Ccal^{(t-1)})\right], \label{eq:waveguide_marginal_gain}\\
\Ccal^{(t)}&=\Ccal^{(t-1)}\cup\{\zb_{j_t}\}.\label{eq:waveguide_codebook_update}
\end{align}
Repeating \eqref{eq:waveguide_marginal_gain}-\eqref{eq:waveguide_codebook_update} until $f(\Ccal^{(Q)})=1$ yields a codebook size satisfying $Q\le\left(1+\ln\frac{1}{w_{\min}}\right)Q_{\scriptscriptstyle\mathrm{opt}}$ \cite{setcover}, where $Q_{\scriptscriptstyle\mathrm{opt}}$ is the optimal cardinality of \eqref{eq:set_cover} and $w_{\min}=\min_{n\in[V]}w_n$.

Re-index the $Q$ retained codewords by $i\in[Q]$ with design centers $\Cb=(\cb_1,\ldots,\cb_Q)$. Greedy selection guarantees $\max_{i\in[Q]}\Gamma_{i,n}\ge\tau_\delta$ for every $n\in[V]$, yet subregions near the threshold may retain thin coverage margins. We therefore allow each $\cb_i$ to move away from the fixed subregion centers $\qb_n$ and resynthesize $\zb_i(\cb_i)$ via Algorithm~\ref{alg:pa_synthesis} to maximize the weighted positive margin:
\begin{align}
F_{\scriptscriptstyle\mathrm{cen}}(\Cb)&=\sum_{n=1}^{V}w_n
\left[\max_{i\in[Q]}m_{i,n}(\cb_i)\right]_+,
\label{eq:center_phi}\\
m_{i,n}(\cb)&=\Gamma_{i,n}(\cb)-\tau_\delta,
\label{eq:center_margin}
\end{align}
where $\Gamma_{i,n}(\cb)$ is evaluated via \eqref{eq:certificate_power} over the fixed subregion $\Vcal_n$ for the configuration $\zb_i(\cb)$ returned by Algorithm~\ref{alg:pa_synthesis} at design center $\cb$. At iteration $t$, each subregion is assigned to its highest-margin codeword, forming the assigned index set $\Vcal_i^{(t)}$ and its covered subset $\Scal_i^{(t)}=\{n\in\Vcal_i^{(t)}:m_{i,n}(\cb_i^{(t)})\ge0\}$. Each center is then updated over the line-search set $\mathcal L_i^{(t)}$ defined in Appendix~\ref{app:center_refinement} without losing certified coverage:
\begin{subequations}
\label{eq:center_linesearch}
\begin{align}
&\cb_i^{(t+1)}\in\argmax_{\cb\in\mathcal L_i^{(t)}}
\sum_{n\in\Vcal_i^{(t)}}w_n[m_{i,n}(\cb)]_+\\
\mathrm{s.t.}\quad&
m_{i,n}(\cb)\ge0,\quad \forall n\in\Scal_i^{(t)}.
\label{eq:center_coverage_constraint}
\end{align}
\end{subequations}
The refined codebook satisfies \eqref{eq:complete_coverage} whenever
\begin{align}
\max_{n\in[V]}\delta_n^{\scriptscriptstyle\mathrm{crt}}&\le\delta,
\label{eq:codebook_coverage_check}\\
\delta_n^{\scriptscriptstyle\mathrm{crt}}
&=-10\log_{10}\left(
\max_{i\in[Q]}\Gamma_{i,n}(\cb_i)
\right).
\label{eq:cell_certified_loss}
\end{align}
Both $\Ccal$ and its spatial responses are stored offline for beam training.

\subsection{Multi-Waveguide Power Coverage and Channel Conditioning}

To serve $U$ users simultaneously across $W$ waveguides, a joint configuration $\Zb_p=(\zb_1^{(p)},\ldots,\zb_W^{(p)})\in\Ccal^W$ assigns one codeword from $\Ccal$ to each waveguide. Effective multi-user transmission requires both sufficient aggregate received power per user and well-conditioned spatial separation across users. Whereas $\Ccal$ is designed over the waveguide-relative domain $\Omega_{\mathrm r}$, multi-user links depend on physical user locations in $\Omega_{xy}$. We therefore partition $\Omega_{xy}$ into $V_{xy}$ user-location subregions $\{\Vcal_n^{xy}\}_{n=1}^{V_{xy}}$ with representative centers $\pb_n^{xy}=[x_n,y_n,0]^T$. A tuple of user-location subregion indices is denoted by $\mathbf n=(n_1,\ldots,n_U)\in[V_{xy}]^U$.

For user $u$, the aggregate received gain under $\Zb_p$ and the pool reference gain sum the per-waveguide powers as
\begin{align}
E_u(\Zb_p;\pb_u)
&=\sum_{w=1}^{W}\Pcal(\zb_w^{(p)};\qb_w(\pb_u)),
\label{eq:aggregate_response}\\
E_{u,0}(\pb_u)
&=\sum_{w=1}^{W}\Pcal_0(\qb_w(\pb_u)).
\label{eq:response_reference}
\end{align}
Let $\Gamma_{p,w,n}$ denote the bound in \eqref{eq:certificate_power} for $\zb_w^{(p)}$, obtained by applying Appendix~\ref{app:cell_coverage} to a rectangle enclosing $\qb_w(\Vcal_n^{xy})=\{\qb_w(x,y):(x,y)\in\Vcal_n^{xy}\}$, with reference point $\qb_w(\pb_n^{xy})$. Denote the minimum and maximum transverse distances between waveguide $w$ and $\Vcal_n^{xy}$ by $\rho_{w,n}^-$ and $\rho_{w,n}^+$. Combining the per-waveguide bounds yields the multi-waveguide power margin on $\Vcal_n^{xy}$:
\begin{equation}
m_{p,n}^{\scriptscriptstyle\mathrm{rsp}}
\!=\!\!N\eta^2\sum_{w=1}^{W}
\left(\!
\frac{[\Gamma_{p,w,n}\!-\!\tau_\delta]_+}{(\rho_{w,n}^+)^2}
\!-\!\frac{[\tau_\delta\!-\!\Gamma_{p,w,n}]_+}{(\rho_{w,n}^-)^2}\!
\right).
\label{eq:response_margin}
\end{equation}
Accordingly, the joint power coverage indicator and the center-based channel conditioning indicator are defined as
\begin{align}
J_{p,\mathbf n}^{\scriptscriptstyle\mathrm{pow}}(\delta)
&=\ind\!\left\{
m_{p,n_u}^{\scriptscriptstyle\mathrm{rsp}}\ge0,\ \forall u\in[U]
\right\},
\label{eq:joint_indicator}
\end{align}

\begin{theorem}[Multi-Waveguide Power Coverage Guarantee]
\label{thm:joint_coverage}
\emph{For a joint configuration $\Zb_p\in\Ccal^W$ and a tuple of user-location subregion indices $\mathbf n$, if $J_{p,\mathbf n}^{\scriptscriptstyle\mathrm{pow}}(\delta)=1$, then for every $(x_u,y_u)\in\Vcal_{n_u}^{xy}$,
\begin{equation}
E_u(\Zb_p;\pb_u)
\ge\tau_\delta\sum_{w=1}^{W}\Pcal^\star(\qb_w(\pb_u))
\ge\tau_\delta E_{u,0}(\pb_u).
\label{eq:joint_coverage}
\end{equation}}
\end{theorem}
Applying the gain bound to $\mathbf q_w(\mathcal V_{n_u}^{xy})$ and using $\rho_{w,n_u}^-\le\rho_w(y_u)\le\rho_{w,n_u}^+$ gives
\begin{align}
&E_u(\Zb_p;\pb_u)-\tau_\delta\sum_{w=1}^{W}\Pcal^\star(\qb_w(\pb_u))\notag\\
&\quad\ge N\eta^2\sum_{w=1}^{W}
\frac{\Gamma_{p,w,n_u}-\tau_\delta}{\rho_w(y_u)^2}
\ge m_{p,n_u}^{\scriptscriptstyle\mathrm{rsp}}\ge0,
\end{align}
where the last inequality follows from $J_{p,\mathbf n}^{\scriptscriptstyle\mathrm{pow}}(\delta)=1$ in \eqref{eq:joint_indicator}, and $\Pcal^\star(\qb)\ge\Pcal_0(\qb)$ yields the second inequality in \eqref{eq:joint_coverage}.

\subsection{Coordinate-Search-Based Joint Configuration Selection}

Evaluating all $Q^W$ combinations in $\Ccal^W$ is prohibitive for multiple waveguides. Because the $w$-th column of $\Gb(\Zb;\Pb)$ in \eqref{eq:effective_channel} depends solely on $\zb_w$, changing one waveguide alters only a single channel column. Exploiting this column structure, we build a compact joint candidate set $\Ccal_W\subseteq\Ccal^W$ via per-waveguide coordinate search.

Each waveguide $w\in[W]$ is supplied with a finite set $\Acal_w$ of available channel columns $\{\mathbf h_{w,q}\in\Cbb^U:q\in\Acal_w\}$, per-entry error variance $\nu$, and codeword map $\iota_w$. In Section~\ref{sec:training}, these columns come from two complementary sources: noisy probing measurements and stored responses at likely user-location tuples. An index vector $\ib=(i_1,\ldots,i_W)\in\Acal_1\times\cdots\times\Acal_W$ defines
\begin{subequations}
\label{eq:joint_candidate_objective}
\begin{align}
\mathbf H(\ib)&=[\mathbf h_{1,i_1},\ldots,\mathbf h_{W,i_W}],
\label{eq:joint_channel_assembly}\\
\Zb(\ib)&=(\zb_{\iota_1(i_1)},\ldots,\zb_{\iota_W(i_W)}).
\label{eq:joint_activation_assembly}
\end{align}
\end{subequations}
To remove mean noise bias from measured columns while preserving positive semidefiniteness, we form the debiased Gram matrix
\begin{equation}
\widehat\Rb(\mathbf H;\nu)=\Pi_{\succeq0}\left(\mathbf H\mathbf H^H-W\nu\Ib_U\right),
\label{eq:debiased_gram}
\end{equation}
where $\Pi_{\succeq0}$ projects onto the positive semidefinite cone. Let $\{\lambda_j\}_{j=1}^{U}$ and $\{e_u\}_{u=1}^{U}$ denote the eigenvalues and diagonal entries of $\widehat\Rb(\mathbf H;\nu)$. We rank candidate joint configurations using the following measure computed from the available channel columns:
\begin{equation}
\begin{aligned}
\Psi(\mathbf H;\nu)
&=\!\!\frac1U\!\!\sum_{j=1}^{U}\log_2\!\left(1\!+\!\frac{\gamma_{\mathrm{tx}}}{U}\lambda_j\right)\! +\!\omega_f\!\min_{u\in[U]}\!\log_2\!\left(1\!+\!\frac{\gamma_{\mathrm{tx}}}{U}\!e_u\!\right)\\
&\quad+\omega_c\log_2\left(1+\frac{\gamma_{\mathrm{tx}}}{U}\min_{j\in[U]}\lambda_j\right),
\end{aligned}
\label{eq:training_objective}
\end{equation}
where $\omega_f,\omega_c\ge0$ weight weakest-user power via $e_u$ and channel separability via $\min_j\lambda_j$ alongside a measure of overall channel quality via $\lambda_j$. The shortlisted candidates are subsequently confirmed with pilots in Section~\ref{sec:training}.

For each initial vector $\ib_p^{(0)}$, coordinate search sets $\ib_p=\ib_p^{(0)}$ and updates one waveguide at a time over the full multi-user matrix:
\begin{equation}
i_{p,w}\in
\argmax_{j\in\Acal_w}\Psi(\mathbf H(\ib_p[w\leftarrow j]);\nu),
\label{eq:joint_block_update}
\end{equation}
where $\ib_p[w\leftarrow j]$ replaces the $w$-th entry of $\ib_p$ with $j$, breaking ties arbitrarily. Algorithm~\ref{alg:codebook} searches both column collections, places location-derived candidates ahead of $\Psi$-ranked measurement-derived candidates, removes duplicates, and truncates the list to the budget $Q_W$.

\begin{algorithm}[t]
\caption{Joint Configuration Construction for the Current Users}
\label{alg:codebook}
\begin{algorithmic}[1]
\Require Codebook $\Ccal$, sweep limit, budget $Q_W$, and column collections with source $s\in\{\mathrm{loc},\mathrm{mea}\}$, indices $\{\Acal_w\}$, variance $\nu$, maps $\{\iota_w\}$, and initial vectors.
\Ensure Joint candidate set $\Ccal_W\subseteq\Ccal^W$.
\STATE Initialize empty lists $L_{\scriptscriptstyle\mathrm{loc}}$ and $L_{\scriptscriptstyle\mathrm{mea}}$.
\FOR{each column collection with source $s$}
\FOR{each initial index vector $\ib_p^{(0)}$}
\STATE Set $\ib_p\leftarrow\ib_p^{(0)}$.
\FOR{each prescribed sweep $r$}
\FOR{$w=1,\ldots,W$}
\STATE Update $i_{p,w}$ via \eqref{eq:joint_block_update} using the current collection.
\ENDFOR
\ENDFOR
\STATE Assemble $\mathbf H(\ib_p)$ and $\Zb(\ib_p)$ via \eqref{eq:joint_candidate_objective}, and append $(\Zb(\ib_p),\Psi(\mathbf H(\ib_p);\nu))$ to $L_s$.
\ENDFOR
\ENDFOR
\STATE Stably sort $L_{\scriptscriptstyle\mathrm{mea}}$ by decreasing $\Psi$.
\STATE Concatenate the configurations in $L_{\scriptscriptstyle\mathrm{loc}}$ followed by those in $L_{\scriptscriptstyle\mathrm{mea}}$ into $L$.
\STATE Remove repeated configurations from $L$, keeping the first occurrence.
\STATE Set $\Ccal_W$ to the first $\min\{Q_W,|L|\}$ entries of $L$.
\RETURN $\Ccal_W$.
\end{algorithmic}
\end{algorithm}

% =====================================================================
\section{Codebook-Based Joint Beam Training}
\label{sec:training}
% =====================================================================

With the shared codebook $\Ccal$ and its stored responses, beam training selects a joint configuration for unknown user locations and channels within $T$ training blocks. Stage~I acquires channel columns and estimates location probabilities. Stage~II supplies measured and stored columns to Algorithm~\ref{alg:codebook} and forms the confirmation shortlist $\Kcal_c$. Stage~III measures the shortlisted candidates to obtain $\widehat\Zb$ and $\widehat\Gb$.

\subsection{Stage I: Adaptive Probing and Location Estimation}

Stage~I aims to acquire informative channel measurements and narrow down user locations. For each user $u\in[U]$, we maintain a location probability distribution $\pi_{u,t}(n)$ over $V_p$ user-location grid points $\pb_n^{\mathrm g}\in\Omega_{xy}\times\{0\}$, $n\in[V_p]$, initialized uniformly as $\pi_{u,0}(n)=1/V_p$. Probing selection, pilot measurement, and probability updating then form a closed feedback loop across blocks $t\in[T_s]$.

In pilot block $t$, waveguide $w$ activates codeword $\zb_{i_{t,w}}\in\Ccal$, forming the joint configuration $\Zb_t=(\zb_{i_{t,1}},\ldots,\zb_{i_{t,W}})$ indexed by $\ib_t=(i_{t,1},\ldots,i_{t,W})$. The $W$ RF chains transmit orthogonal pilot sequences $\Sb\in\Cbb^{W\times\tau_p}$ with total block energy $E_{\mathrm{tr}}$, where $\Sb\Sb^H=\tau_p\Ib_W$ and $\tau_p\ge W$. The received pilot matrix and despread channel estimate are
\begin{align}
\Yb_t&=\sqrt{\frac{E_{\mathrm{tr}}}{W\tau_p}}\,
\Gb(\Zb_t;\Pb)\Sb+\Nb_t,
\label{eq:parallel_pilot}\\
\widehat\Gb_t&=
\sqrt{\frac{W\tau_p}{E_{\mathrm{tr}}}}\,
\frac{\Yb_t\Sb^H}{\tau_p}
=\Gb(\Zb_t;\Pb)+\widetilde\Nb_t,
\label{eq:parallel_despread}
\end{align}
where $\Nb_t$ has independent $\CN(0,\sigma_p^2)$ entries, each entry of $\widetilde\Nb_t$ has variance $\sigma_e^2=W\sigma_p^2/E_{\mathrm{tr}}$, and $\gamma_{\mathrm{tr}}=E_{\mathrm{tr}}/\sigma_p^2$ is the training SNR based on the total pilot-block energy. To choose the next unprobed codeword on waveguide $w$, the probing criterion adapts to $\gamma_{\mathrm{tr}}$. At low training SNR, probing prioritizes codewords with wide strong-response regions and high average power through a baseline score $B_i$ and greedy rank preference. As $\gamma_{\mathrm{tr}}$ increases, probing shifts toward codewords with strong posterior response variation $V_i$ to distinguish competing user-location hypotheses. Across all SNRs, a diversity score $D_i$ penalizes correlation with already tested codewords in $\Mcal_{t,w}$ to avoid redundant probes. Let $h_{i,w}(\pb)$ denote the channel in \eqref{eq:scalar_channel} stored for codeword $i$ on waveguide $w$ at $\pb$, with shorthand $h_i(n)=h_{i,w}(\pb_n^{\mathrm g})$ collected in $\hb_i\in\Cbb^{V_p}$. The probed index set starts at $\Mcal_{0,w}=\varnothing$ and is updated as $\Mcal_{t,w}=\Mcal_{t-1,w}\cup\{i_{t,w}\}$. The next codeword is chosen as
\begin{align}
&i_{t+1,w}\in\argmax_{i\in[Q]\setminus\Mcal_{t,w}}S_i,
\label{eq:adaptive_measurement_rule}\\
&S_i\!\!= \!\chi(V_i\! +\! \omega_{\scriptscriptstyle\mathrm{div}}D_i) \!
+\! (1\!-\!\chi)\!\left(\!B_i\! +\! \frac{\omega_{\scriptscriptstyle\mathrm{rnk}}}{1\!+\!r_i}\!\right)
\!+\! \omega_{\scriptscriptstyle\mathrm{bas}}D_i,
\label{eq:adaptive_measurement_criterion}
\end{align}
where $\chi=\gamma_{\mathrm{tr}}/(\gamma_{\mathrm{tr}}+\gamma_{\scriptscriptstyle\mathrm{mix}})\in[0,1]$ mixes the two regimes around the linear reference SNR $\gamma_{\scriptscriptstyle\mathrm{mix}}>0$, $r_i$ is the greedy selection rank of codeword $i$, and $\omega_{\scriptscriptstyle\mathrm{div}},\omega_{\scriptscriptstyle\mathrm{rnk}},\omega_{\scriptscriptstyle\mathrm{bas}}\ge0$ weight diversity and rank preferences. With the dependence on $t$ and $w$ omitted from the component names, the three scoring components are
\begin{align}
V_i&=\frac1U\sum_{u=1}^{U}
\frac{\sum_{n=1}^{V_p}\pi_{u,t}(n)
|h_i(n)-\bar h_{i,u}|^2}
{\sum_{n=1}^{V_p}\pi_{u,t}(n)|h_i(n)|^2+\epsilon_0},
\label{eq:posterior_variance}\\
D_i&=
\min_{j\in\Mcal_{t,w}}
\left(1-
\frac{|\widetilde\hb_i^H\widetilde\hb_j|}
{\|\widetilde\hb_i\|_2\|\widetilde\hb_j\|_2+\epsilon_0}
\right),
\label{eq:measurement_correlation}\\
B_i&=\!\omega_{\scriptscriptstyle\mathrm{wid}}\frac{A_i}{\max_{j\in[Q]} \!A_j\!+\!\epsilon_0} \!+\!(1\!-\!\omega_{\scriptscriptstyle\mathrm{wid}})\frac{E_i}{\max_{j\in[Q]} E_j+\epsilon_0},
\label{eq:stored_response_measure}
\end{align}
where $\epsilon_0>0$ ensures numerical stability, $\bar h_{i,u}=\sum_{n=1}^{V_p}\pi_{u,t}(n)h_i(n)$ is the posterior mean response, and $\widetilde\hb_i=\hb_i-(V_p^{-1}\mathbf{1}^T\hb_i)\mathbf{1}$ is the centered response vector with all-ones vector $\mathbf{1}$. For $\Mcal_{t,w}=\varnothing$, we set $D_i=1$. The width weight $\omega_{\scriptscriptstyle\mathrm{wid}}\in[0,1]$ balances the strong-gain area fraction $A_i=V_p^{-1}\sum_{n=1}^{V_p}\ind\{|h_i(n)|^2\ge\tau_\delta\max_{m\in[V_p]}|h_i(m)|^2\}$ against the mean power $E_i=V_p^{-1}\sum_{n=1}^{V_p}|h_i(n)|^2$.

After receiving $\widehat\Gb_t$, we update $\{\pi_{u,t}(n)\}$ by matching accumulated measurements against stored responses. Omitting the block index $t$, let $\hb_n^{\mathrm g}=\operatorname{vec}\big([h_{i_{b,w},w}(\pb_n^{\mathrm g})]_{b\in[t],w\in[W]}\big)$ and $\widehat\hb_u=\operatorname{vec}\big([[\widehat\Gb_b]_{u,w}]_{b\in[t],w\in[W]}\big)$ stack the model responses at $\pb_n^{\mathrm g}$ and the measurements for user $u$ in the same order. Because an unknown complex path gain $\alpha$ scales the measured channel, we estimate it by Euclidean projection onto $\Acal_\alpha=\{\alpha\in\Cbb:\alpha_{\min}\le|\alpha|\le\alpha_{\max}\}$ and compute the normalized matching residual:
\begin{align}
\widehat\alpha_{u,n}
&=\Pi_{\Acal_\alpha}\!\left(
\frac{(\hb_n^{\mathrm g})^H\widehat\hb_u}
{\|\hb_n^{\mathrm g}\|_2^2+\epsilon_0}
\right),
\label{eq:calibrated_scale}\\
d_{u,n}&=\frac{
\|\widehat\hb_u-\widehat\alpha_{u,n}\hb_n^{\mathrm g}\|_2^2}
{\|\widehat\hb_u\|_2^2+\epsilon_0}.
\label{eq:state_residual_new}
\end{align}
Let $\Ncal_u$ collect the $K_g\in[V_p]$ search points with the smallest residuals for user $u$. Setting $\Delta d_{u,n}=d_{u,n}-\min_{m\in[V_p]}d_{u,m}$ and $s_u=\max\{\epsilon_d,\operatorname{median}_{n\in\Ncal_u}\Delta d_{u,n}\}$ with lower bound $\epsilon_d>0$, the location probabilities are updated as
\begin{equation}
\pi_{u,t}(n)=
\frac{\exp(-\Delta d_{u,n}/s_u)}
{\sum_{m\in\Ncal_u}\exp(-\Delta d_{u,m}/s_u)},
\quad n\in\Ncal_u.
\label{eq:state_posterior_new}
\end{equation}
For $n\notin\Ncal_u$, $\pi_{u,t}(n)=0$.
\subsection{Stage II: Joint Search and Off-Grid Refinement}
Stage~II constructs a compact set of joint configurations from the Stage~I outputs and optionally refines their PA positions beyond the discrete codebook. Since the $w$-th column of $\Gb(\Zb;\Pb)$ depends only on waveguide $w$'s configuration, measured columns from different probing blocks $\tb\!=\!(t_1,\ldots,t_W)\!\in\![T_s]^W$ can be recombined without additional pilots:
\begin{equation}
\widehat\Gb(\tb)=
\left[[\widehat\Gb_{t_1}]_{:,1},\ldots,
[\widehat\Gb_{t_W}]_{:,W}\right],
\label{eq:measured_column_assembly}
\end{equation}
where $[\widehat\Gb_t]_{:,w}$ denotes the $w$-th column of $\widehat\Gb_t$. These measured columns are supplied to Algorithm~\ref{alg:codebook} with $\Acal_w=[T_s]$, $\mathbf h_{w,t}=[\widehat\Gb_t]_{:,w}$, $\iota_w(t)=i_{t,w}$, and $\nu=\sigma_e^2$, initialized from the strongest measured configurations.

To include unprobed codewords in the search, two user-location tuples are formed from \eqref{eq:state_posterior_new} after block $T_s$. The first tuple uses each user's highest-weight search point and the second uses each user's second-highest-weight point. For either tuple $\widehat\Pb=(\widehat\pb_1,\ldots,\widehat\pb_U)$, the corresponding stored channel columns are
\begin{equation}
[\mathbf h_{w,i}]_u=h_{i,w}(\widehat\pb_u),
\label{eq:model_channel_assembly}
\end{equation}
where $\Acal_w\subseteq[Q]$ is selected based on the mean and minimum logarithmic user gains. This model-based collection is supplied with $\iota_w(i)=i$ and $\nu=0$. Algorithm~\ref{alg:codebook} processes the measured collection and the collection for each location tuple separately, merges the resulting configurations without duplicates, and retains at most $Q_W$ candidates in $\Ccal_W$, where $1\le Q_W\le\min\{4,T-T_s\}$.

When off-grid refinement is activated under $\gamma_{\mathrm{tr}}>\gamma_{\scriptscriptstyle\mathrm{off}}$ and $T-T_s\ge T_{\min}^{\scriptscriptstyle\mathrm{off}}=9$. Each candidate in $\Ccal_W$ is measured once. The baseline $\Zb^0$ is selected by maximizing $\mathcal J(\Rb_\epsilon(\mathbf H_p;\sigma_e^2))$ over the candidates, where $\mathcal J$ is defined in \eqref{eq:offgrid_rate_measure}, $\mathbf H_p$ is the measured channel of candidate $p$, and $\Rb_\epsilon(\mathbf H;\nu)=\Pi_{\succeq\epsilon\Ib_U}(\widehat\Rb(\mathbf H;\nu))$ with $\epsilon=\max\{\zeta_{\scriptscriptstyle\mathrm{reg}}W\nu,\epsilon_{\min}\}$. To reduce overhead, the PAs on waveguide $w$ are partitioned into $\min\{M_w,4,T-T_s-T_b-3\}$ contiguous groups $\{\Gcal_{w,\ell}\}$ whose members move together. One group per waveguide is probed in parallel in each of the $T_g$ blocks, where $T_g$ is the largest group count and $t_{w,\ell}$ counts the blocks for group $\Gcal_{w,\ell}$. Its contribution to the baseline is
\begin{equation}
\vb_{w,\ell}
=\sqrt{\frac{M_{w,\ell}}{M_w}}\,
\overline{\widehat{\gb}}_{w,\ell},
\label{eq:offgrid_group_contribution}
\end{equation}
where $M_{w,\ell}=|\Gcal_{w,\ell}|$ and each entry of $\overline{\widehat{\gb}}_{w,\ell}$ has variance $\sigma_e^2/t_{w,\ell}$.

To predict the channel after a group translation without additional pilots, we approximate the change in the combined response of each PA group by a single complex factor. At each estimated user location, this factor is computed as the arithmetic mean of the individual PA response ratios. Applying these factors to the measured group contributions yields the following channel prediction and propagated measurement-noise variance:
\begin{align}
\widetilde{\gb}_w
&=\sum_{\ell}\bb_{w,\ell}\odot\vb_{w,\ell},
\label{eq:offgrid_predicted_channel}\\
\boldsymbol\nu_w
&=\sigma_e^2\sum_{\ell}
\frac{M_{w,\ell}}{M_wt_{w,\ell}}
|\bb_{w,\ell}|^2,
\label{eq:offgrid_predicted_variance}\\
[\bb_{w,\ell}]_u
&=\frac{1}{M_{w,\ell}}\sum_{g\in\Gcal_{w,\ell}}
\frac{a(\xi'_{w,g};\widehat\qb_{w,u})}
{a(\xi^0_{w,g};\widehat\qb_{w,u})},
\label{eq:offgrid_translation_weight}
\end{align}
where $\widehat\qb_{w,u}=\qb_w(\widehat\pb_u)$ uses the first location tuple, $\bb_{w,\ell}$ collects the translation factors, and $\odot$ and $|\cdot|^2$ are applied elementwise.

Given $\widetilde\Gb=[\widetilde{\gb}_1,\ldots,\widetilde{\gb}_W]$, the noise-corrected regularized Gram matrix is
\begin{equation}
\Rb
=\Pi_{\succeq\epsilon\Ib_U}
\left(\widetilde\Gb\widetilde\Gb^H-\Db\right),
\label{eq:offgrid_predicted_gram}
\end{equation}
where $\Db=\diag\left(\sum_{w=1}^{W}\boldsymbol\nu_w\right), \epsilon =\max\{\zeta_{\scriptscriptstyle\mathrm{reg}}\|\Db\|_2,\epsilon_{\min}\}.$ $\Pi_{\succeq\epsilon\Ib_U}$ projects onto Hermitian matrices with eigenvalues no smaller than $\epsilon$, while $\zeta_{\scriptscriptstyle\mathrm{reg}},\epsilon_{\min}>0$ control the regularization level and ensure invertibility.

Group translations are optimized via coordinate search to maximize the multi-user rate metric
\begin{subequations}
\label{eq:offgrid_rate_measure}
\begin{align}
\mathcal J(\Rb)
&=\frac1U\sum_{u=1}^{U}\widehat R_u(\Rb)
+\omega_{\scriptscriptstyle\mathrm{off}}\min_{u\in[U]}\widehat R_u(\Rb),
\label{eq:offgrid_joint_rate}\\
\widehat R_u(\Rb)
&=\log_2\left(1+\frac{\gamma_{\mathrm{tx}}}
{U[\Rb^{-1}]_{u,u}}\right),
\label{eq:offgrid_user_rate}
\end{align}
\end{subequations}
where $\omega_{\scriptscriptstyle\mathrm{off}}\ge0$ weights the weakest user's estimated rate. Finally, $\Zb^0$ and up to two refined configurations form the confirmation shortlist $\Kcal_c$. If the off-grid step is skipped, $T_b=T_g=0$ and $\Kcal_c=\Ccal_W$.

\subsection{Stage III: Candidate Confirmation and Channel Estimation}
Next, direct measurement is required for reliable final selection and channel estimation. Stage~III allocates the remaining $T_c=T-T_s-T_b-T_g$ training blocks across $\Kcal_c$ such that $t_p\ge1$ and $\sum_{p=1}^{|\Kcal_c|}t_p=T_c$. Averaging $t_p$ independent blocks for candidate $\Zb_p^{\mathrm c}\in\Kcal_c$ lowers the per-entry noise variance to $\nu=\sigma_e^2/t_p$ and produces
\begin{equation}
\overline\Gb_p=\frac1{t_p}\sum_{b=1}^{t_p}\widehat\Gb_{p,b}.
\label{eq:confirmed_channel_average}
\end{equation}
Using the directly measured channels, the final configuration $\widehat\Zb=\Zb_{\widehat p}^{\mathrm c}$ is selected via
\begin{equation}
\widehat p\in
\begin{cases}
\argmax_{p\in[|\Kcal_c|]}\Psi(\overline\Gb_p;\sigma_e^2/t_p),
&\text{on-grid},\\
\argmax_{p\in[|\Kcal_c|]}
\mathcal J\big(\Rb_\epsilon(\overline\Gb_p;\sigma_e^2/t_p)\big),
&\text{off-grid},
\end{cases}
\label{eq:joint_confirmation_new}
\end{equation}
applying $\Psi$ for on-grid selection and the regularized rate metric $\mathcal J$ for off-grid selection. The corresponding averaged channel $\widehat\Gb=\overline\Gb_{\widehat p}$ is returned for transmission. Algorithm~\ref{alg:joint_training} summarizes the complete three-stage procedure.

\begin{algorithm}[t]
\caption{Beam Training with the Proposed Codebook}
\label{alg:joint_training}
\begin{algorithmic}[1]
\Require Codebook $\Ccal$, stored responses, total budget $T$, probing budget $1\le T_s<T$ with $T_s\le Q$, candidate budget $1\le Q_W\le\min\{4,T-T_s\}$, probing weights, and thresholds $\gamma_{\scriptscriptstyle\mathrm{off}}$ and $T_{\min}^{\scriptscriptstyle\mathrm{off}}$.
\Ensure Selected configuration $\widehat\Zb$ and channel $\widehat\Gb$.
\STATE Initialize $\pi_{u,0}(n)=1/V_p$ for all $u\in[U]$ and $n\in[V_p]$. Set $\Mcal_{0,w}=\varnothing$ for all $w\in[W]$.
\FOR{$t=1,\ldots,T_s$}
\STATE Select $\ib_t$ via \eqref{eq:adaptive_measurement_rule}-\eqref{eq:adaptive_measurement_criterion} using information through block $t-1$.
\STATE Measure $\widehat\Gb_t$ via \eqref{eq:parallel_despread}.
\STATE Update $\Mcal_{t,w}\leftarrow\Mcal_{t-1,w}\cup\{i_{t,w}\}$ for each $w\in[W]$.
\STATE Update $\pi_{u,t}(n)$ via \eqref{eq:calibrated_scale}-\eqref{eq:state_posterior_new}.
\ENDFOR
\STATE Form measured and stored column collections via \eqref{eq:measured_column_assembly} and \eqref{eq:model_channel_assembly}.
\STATE Supply both collections to Algorithm~\ref{alg:codebook} to obtain $\Ccal_W$.
\STATE Initialize $T_b=T_g=0$.
\IF{off-grid training is enabled, $\gamma_{\mathrm{tr}}>\gamma_{\scriptscriptstyle\mathrm{off}}$, and $T-T_s\ge T_{\min}^{\scriptscriptstyle\mathrm{off}}$}
\STATE Measure each candidate in $\Ccal_W$ once ($T_b=|\Ccal_W|$) and select $\Zb^0$ with $\mathcal J(\Rb_\epsilon(\mathbf H;\sigma_e^2))$.
\STATE Probe PA groups in parallel over $T_g$ blocks as specified in Stage~II and compute $\vb_{w,\ell}$ via \eqref{eq:offgrid_group_contribution}.
\STATE Optimize group translations via \eqref{eq:offgrid_predicted_channel}-\eqref{eq:offgrid_rate_measure}.
\STATE Form $\Kcal_c$ from $\Zb^0$ and up to two refined candidates.
\ELSE
\STATE Set $\Kcal_c=\Ccal_W$.
\ENDIF
\STATE Allocate $T_c=T-T_s-T_b-T_g$ blocks across $\Kcal_c$ with $t_p\ge1$.
\STATE Select $\widehat\Zb$ and $\widehat\Gb$ via \eqref{eq:confirmed_channel_average}-\eqref{eq:joint_confirmation_new}.
\RETURN $\widehat\Zb$ and $\widehat\Gb$.
\end{algorithmic}
\end{algorithm}

\section{Numerical Results}
\label{sec:results}
% =============================================================
In this section, we evaluate the performance of the proposed reconfigurable-aperture codebooks and joint beam training. The carrier frequency is $28$ GHz. Each waveguide has length $L=10$ m and height $H=2$ m. The effective refractive index is $\neff=1.4$, and $\eta$ is normalized to one. At most $N=36$ PAs are activated on each waveguide, with $d_{\min}=\lambda/2$. Users are drawn uniformly from a $10\,\mathrm m\times10\,\mathrm m$ service region. A seven-waveguide deployment spans $y=1$ m to $y=9$ m. Smaller deployments use its first $W$ waveguides. Both single-waveguide single-user (SWSU) and MWMU settings are considered. Every rate point was averaged over the same $100$ random user tuples for all methods. Unless otherwise stated, $T=48$ and $\gamma_{\mathrm{tr}}=\gamma_{\mathrm{tx}}=30$~dB. At this operating point, the proposed methods use $35$ blocks for adaptive probing. 

The probing weights are $\omega_{\scriptscriptstyle\mathrm{div}}=0.35$, $\omega_{\scriptscriptstyle\mathrm{rnk}}=0.10$, $\omega_{\scriptscriptstyle\mathrm{bas}}=0.05$, and $\omega_{\scriptscriptstyle\mathrm{wid}}=0.6$, with $\gamma_{\scriptscriptstyle\mathrm{mix}}=10$. Joint search uses $\omega_f=0.20$ and $\omega_c=0.15$. The labels below identify the codebook and training procedure used in each curve.
\begin{itemize}

\item \emph{Perfect CSI (P-CSI):} uses exact user locations and noiseless channels in \eqref{eq:true_joint_reference} as the performance benchmark. 

\item \emph{3SBT} is the conventional PASS beam-training method in \cite{pass_train}. The original search procedure is adapted to the considered multi-user setting and common training budget by replacing the separate searches with joint probing and confirmation based on \eqref{eq:true_joint_utility}.

\item \emph{Near-field codebook (Fix-NF)} directly applies a conventional near-field codebook in \cite{neartrain1} to PASS with $N=36$ PAs.

\item \emph{Independent codeword selection (Prop.-IND)} uses the proposed codebook and codeword measurements but selects the waveguide codewords independently.

\item \emph{Proposed on-grid training (Prop.-ON)} uses the reconfigurable-aperture codebook and adaptive codeword measurements. It does not apply continuous position refinement.

\item \emph{Proposed off-grid training (Prop.-OFF)} adds off-grid PA-position refinement to Prop.-ON. 

\end{itemize}

\begin{figure}[t]
\centering
\includegraphics[width=0.8\columnwidth]{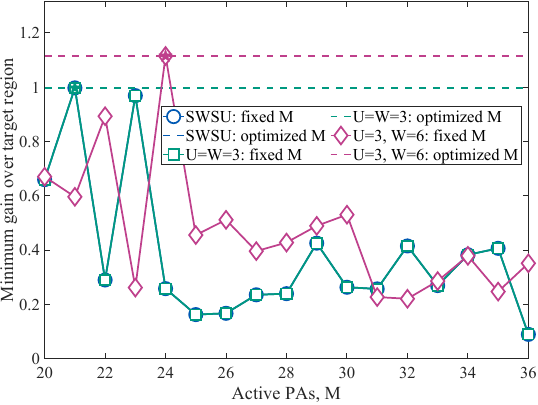}
\caption{Impact of number of active PAs $M$ on the minimum gain over the target region under different $(W,U)$ deployments. ``fix'' uses a fixed number of active PAs, whereas ``opt'' uses the optimized number of active PAs from Algorithm~\ref{alg:pa_synthesis}.}
\label{fig:pa_selection}
\end{figure}

\begin{figure}[t]
\centering
\includegraphics[width=0.8\columnwidth]{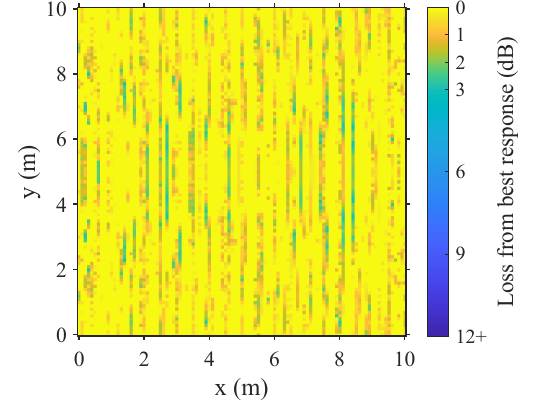}
\caption{Sampled response of the reconfigurable-aperture codebook relative to the candidate-pool maximum.}
\label{fig:swsu_energy_coverage}
\end{figure}

\begin{figure}[t]
\centering
\includegraphics[width=0.8\columnwidth]{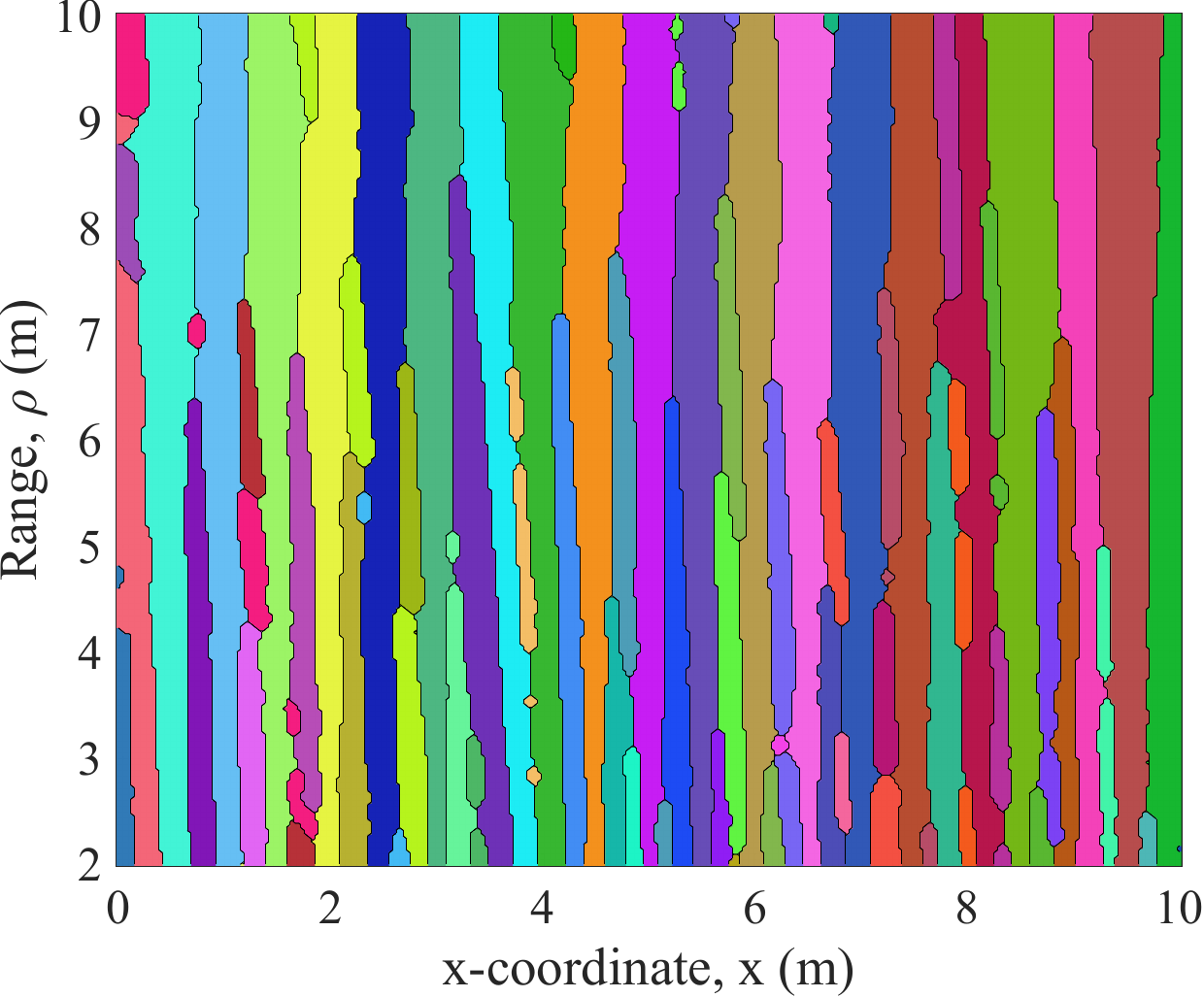}
\caption{Codeword assignment over the $(x,\rho)$ domain. Each color denotes one codeword.}
\label{fig:swsu_codeword_regions}
\end{figure}

\begin{figure}[t]
\centering
\includegraphics[width=0.8\columnwidth]{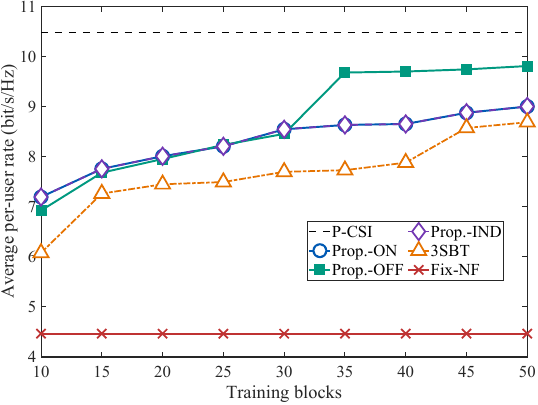}
\caption{Average SWSU rate versus training blocks.}
\label{fig:swsu_rate_pilots}
\end{figure}

\begin{figure}[t]
\centering
\includegraphics[width=0.8\columnwidth]{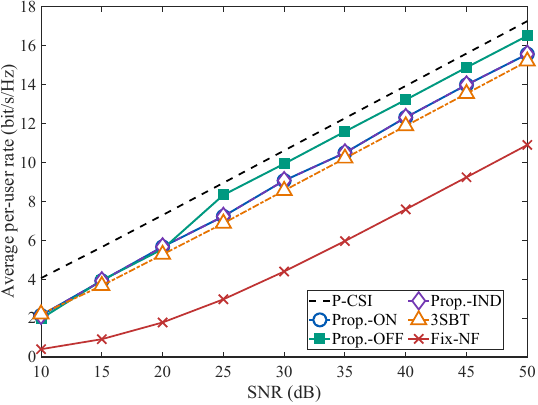}
\caption{Average SWSU rate versus SNR.}
\label{fig:swsu_rate_snr}
\end{figure}

\begin{figure}[t]
\centering
\includegraphics[width=0.8\columnwidth]{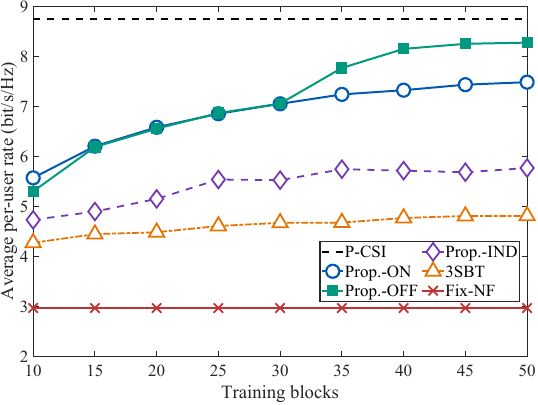}
\caption{Average per-user rate versus training blocks for $U=W=3$.}
\label{fig:mwmu_rate_pilots}
\end{figure}

\begin{figure}[t]
\centering
\includegraphics[width=0.8\columnwidth]{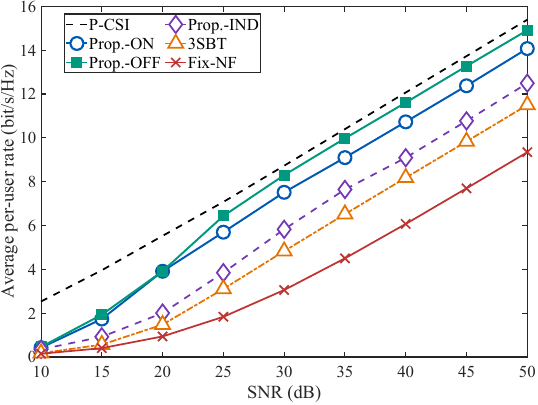}
\caption{Average per-user rate versus SNR for $U=W=3$.}
\label{fig:mwmu_rate_snr}
\end{figure}

\begin{figure}[t]
\centering
\includegraphics[width=0.8\columnwidth]{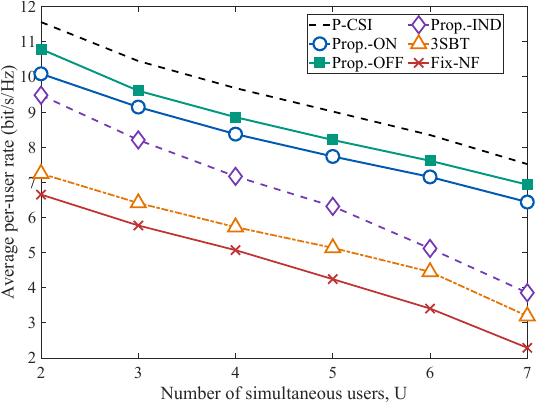}
\caption{Average per-user rate versus number of users.}
\label{fig:mwmu_users_w7}
\end{figure}

\begin{figure}[t]
\centering
\includegraphics[width=0.8\columnwidth]{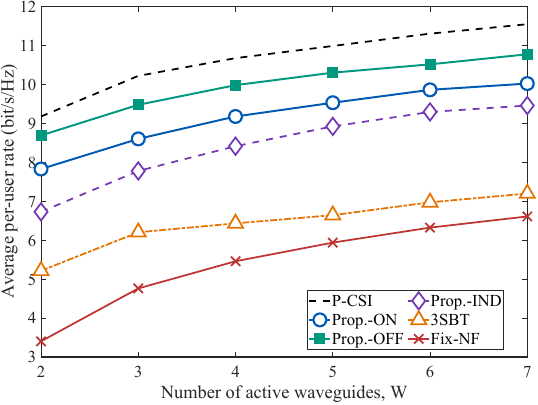}
\caption{Average per-user rate versus number of waveguides.}
\label{fig:mwmu_waveguides_u2}
\end{figure}

\subsection{Codebook Coverage}

Fig.~\ref{fig:pa_selection} evaluates the impact of the number of active PAs $M$ across three $(W,U)$ deployments. As predicted by \eqref{eq:cardinality_difference}, the optimal $M$ varies across deployments. In all cases, Algorithm~\ref{alg:pa_synthesis} matches the best value obtained by sweeping the tested PA counts, confirming that our joint search accurately optimizes the number of active PAs for each codeword.

The refined $138$-codeword codebook satisfies \eqref{eq:codebook_coverage_check} with $\delta=3$~dB on every subregion. By Lemma~\ref{lem:cell_coverage}, the entire service region is therefore covered within $3$~dB of the optimal feasible PA response. Fig.~\ref{fig:swsu_energy_coverage} shows the sampled response relative to the candidate-pool maximum $\Pcal_0(\qb)$. The near-$0$~dB values over most of the grid show that high-response candidates are retained after pruning. Fig.~\ref{fig:swsu_codeword_regions} further illustrates the assignment of covered locations to codewords in the $(x,\rho)$ domain. This coverage pattern differs fundamentally from the uniform partitions of the angle–distance domain, explaining why our spatial-coverage-based codebook is much better matched to PASS propagation than conventional fixed-array codebooks.

\subsection{Single-User Training}
The single-user setting first evaluates the codebook without introducing multi-waveguide gains. Fig.~\ref{fig:swsu_rate_pilots} compares the SWSU achievable rate versus the number of training blocks. Guided by the stored codebook coverage map, Prop.-ON and Prop.-OFF converge rapidly within $10$-$20$ training blocks and consistently outperform 3SBT and Fix-NF. Meanwhile, Fix-NF suffers a persistent $4$-$5$~bit/s/Hz loss across all pilot budgets because additional pilots cannot overcome the beam-shape mismatch of a fixed-aperture codebook. Fig.~\ref{fig:swsu_rate_snr} evaluates the SWSU rate versus transmit SNR. Prop.-ON and Prop.-OFF closely track P-CSI over the entire SNR range, while their advantage over Fix-NF grows with SNR. Moreover, Prop.-OFF provides an additional rate gain over Prop.-ON, confirming the benefit of continuous off-grid refinement at moderate-to-high SNR.

\subsection{Joint Training}

For $U=W=3$, Fig.~\ref{fig:mwmu_rate_pilots} evaluates whether the codebook can support joint beam training for multiple waveguides and users. The gap between Prop.-ON and Prop.-IND increases from $0.8$ to $1.7$~bit/s/Hz as the training budget grows from $10$ to $50$ blocks. Since both methods use the same codebook, this difference isolates the benefit of joint codeword selection. The result shows that the codebook eliminates inefficient candidates, while joint selection further avoids combinations that are individually strong but jointly unfavorable for multi-user transmission. Fig.~\ref{fig:mwmu_rate_snr} examines the robustness of joint training over SNR. Prop.-ON remains close to the P-CSI benchmark over the whole range, whereas 3SBT and Fix-NF stay substantially lower over a wider SNR region. The proposed method therefore retains its advantage even when the available signal power increases.

\subsection{Scalability}

The scalability of joint selection is examined with respect to both the number of users and waveguides. Fig.~\ref{fig:mwmu_users_w7} first tests how joint selection performs as the user count increases. The average per-user rate decreases with the user count for every method, but the gap between Prop.-ON and Prop.-IND widens from $0.7$~bit/s/Hz at $U=2$ to $2.5$~bit/s/Hz at $U=7$. Since their codebook and measurements are otherwise identical, this widening gap indicates that joint selection contributes increasingly to the rate as user interactions grow stronger. Fig.~\ref{fig:mwmu_waveguides_u2} then examines codebook reuse across waveguides at $U=2$. As $W$ grows from $2$ to $7$, the gap between Prop.-ON and each baseline changes by no more than $0.2$~bit/s/Hz, indicating that the added waveguides improve all methods by a similar amount without closing the performance gap. This confirms that the single-waveguide codebook can be reused for joint construction as the deployment grows, without restarting a continuous PA-placement search for each deployment size.

% =====================================================================
\section{Conclusion}
\label{sec:conclusion}
% =====================================================================

This paper proposed a reconfigurable-aperture codebook and joint beam training framework for multi-waveguide PASS. By exploiting phase-aligned PA parameterization and an analytical bound that certifies coverage over each subregion, the proposed method constructs a compact codebook with guaranteed spatial coverage and reuses its coverage information for three-stage beam training, including adaptive probing, joint configuration search, and off-grid PA refinement. Numerical results demonstrated near-optimal coverage and significant rate gains over existing PASS training methods and conventional near-field codebooks across different user and waveguide deployments.
\appendices

\section{Proof of Lemma~\ref{lem:cell_coverage}}
\label{app:cell_coverage}
Fix candidate $\zb_j$ and subregion $\Vcal_n$ with center $\qb_n=[x_n,\rho_n]^T$. In this proof only, write $M=M_j$, $\xi_g=\xi_{j,g}$, $\beta_g=\beta_{j,g,n}$, and $\Theta_g=\Theta_{j,g,n}$. Enclose $\Vcal_n$ in the rectangle $\Rcal_n=[x_n-h_x,x_n+h_x]\times[\rho^-,\rho^+]$, where $h_x$ is the longitudinal half-width and $0<\rho^-\le\rho^+$ bound the transverse coordinate throughout $\Vcal_n$. Set $h_\rho=\max\{\rho_n-\rho^-,\rho^+-\rho_n\}$. Let $\xi_0=(\xi_1+\xi_M)/2$ denote the aperture midpoint, and write $r_g(\qb)=r(\xi_g;\qb)$ for $g=0,\ldots,M$. With $\phi_{\scriptscriptstyle\mathrm{ref}}$ a common reference phase for the given candidate and subregion, let $\vartheta_g(\qb)$ be the residual of $\phi(\xi_g;\qb)-\phi_{\scriptscriptstyle\mathrm{ref}}-k[r_0(\qb)-r_0(\qb_n)]$ wrapped to $[-\pi,\pi)$. The bounds in \eqref{eq:certificate_power} are
\begin{align}
\Theta_g\!\!=\!|\vartheta_g(\qb_n)|\!+\!\!k|\xi_g\!-\!\xi_0|
\left(\frac{h_x}{\rho^-}
\!+\!\frac{h_\rho\max\{d_g,d_0\}}{(\rho^-)^2}\right),
\label{eq:certificate_phase}
\end{align}
where $d_g=|x_n-\xi_g|+h_x$ for $g=0,\ldots,M$, $\beta_g=\frac{\rho^-}{\sqrt{(\rho^-)^2+d_g^2}}$ and $|\vartheta_g(\qb_n)|$ is the absolute phase residual at the subregion center.

Applying the mean value theorem to $r_g(\qb)-r_0(\qb)$ over $\qb\in\Rcal_n$ yields $|\partial_x(r_g-r_0)|\le|\xi_g-\xi_0|/\rho^-$ and $|\partial_\rho(r_g-r_0)|\le|\xi_g-\xi_0|\max\{d_g,d_0\}/(\rho^-)^2$. Integrating from $\qb_n$ to any $\qb\in\Rcal_n$ and factoring out the common phase $\phi_{\scriptscriptstyle\mathrm{ref}}+k[r_0(\qb)-r_0(\qb_n)]$ gives $|\vartheta_g(\qb)|\le\Theta_g$. The normalized amplitude satisfies $\rho/r_g(\qb)\ge\beta_g$. Thus, when $\max_{g\in[M]}\Theta_g\le\pi/2$, substituting these bounds into \eqref{eq:single_link_boundary} gives
\begin{align}
\Pcal(\zb_j;\qb)&\ge\frac{\eta^2}{M\rho^2}
\left(\sum_{g=1}^{M}\frac{\rho}{r_g(\qb)}\cos\vartheta_g(\qb)\right)^2\notag\\
&\ge\frac{\eta^2}{M\rho^2}
\left(\sum_{g=1}^{M}\beta_g\cos\Theta_g\right)^2
=\Gamma_{j,n}\frac{N\eta^2}{\rho^2}.
\label{eq:certificate_response_lower}
\end{align}
Furthermore, for any $\zb_w=(M_w,\boldsymbol\xi_w)\in\Fcal_N$ with $M_w\le N$, the triangle inequality implies
\begin{equation}
\Pcal(\zb_w;\qb)\le\frac{\eta^2}{M_w}
\left(\sum_{g=1}^{M_w}\frac{1}{r(\xi_{w,g};\qb)}\right)^2
\le\frac{N\eta^2}{\rho^2},
\label{eq:certificate_reference_upper}
\end{equation}
and hence $\Pcal_0(\qb)\le \Pcal^\star(\qb)\le N\eta^2/\rho^2$. Combining \eqref{eq:certificate_response_lower} and \eqref{eq:certificate_reference_upper} with $\Gamma_{j,n}\ge\tau_\delta$ establishes \eqref{eq:cell_coverage}.

\section{Design-Center Refinement}
\label{app:center_refinement}
At iteration $t$, each subregion $\mathcal V_n$ is assigned to one codeword $i\in\argmax_{i'\in[Q]}m_{i',n}(\cb_{i'}^{(t)})$. The index set $\Vcal_i^{(t)}$ collects the indices of the assigned subregions to codeword $i$. For each nonempty $\Vcal_i^{(t)}$, the line-search candidate set in \eqref{eq:center_linesearch} is constructed along the segment toward the gain-weighted centroid $\widetilde{\cb}_i^{(t)}$:
\begin{align}
\mathcal L_i^{(t)}
&=\left\{\operatorname{proj}_{\Omega_{\mathrm r}}\!\left(
(1-\tau)\cb_i^{(t)}+\tau\widetilde\cb_i^{(t)}\right):
\tau\in\mathcal T\right\},
\label{eq:center_segment}\\
\widetilde{\cb}_i^{(t)}&=
\frac{\sum_{n\in\Vcal_i^{(t)}}w_n \Pcal_{i,n}^{(t)}\qb_n}
{\sum_{n\in\Vcal_i^{(t)}}w_n \Pcal_{i,n}^{(t)}},\\
\Pcal_{i,n}^{(t)}&=\Pcal(\zb_i(\cb_i^{(t)});\qb_n),
\label{eq:center_update}
\end{align}
where $\operatorname{proj}_{\Omega_{\mathrm r}}$ denotes Euclidean projection onto $\Omega_{\mathrm r}$ and $\mathcal T\subset[0,1]$ is a finite step-size grid containing $0$. Because $0\in\mathcal T$, the current center $\cb_i^{(t)}$ is always feasible for \eqref{eq:center_linesearch}. Thus, updating $\cb_i^{(t+1)}$ via \eqref{eq:center_linesearch} monotonically non-decreases the total positive coverage margin $F_{\scriptscriptstyle\mathrm{cen}}(\Cb)$ in \eqref{eq:center_phi}, while the constraint $m_{i,n}(\cb)\ge0$ for all $n\in\Scal_i^{(t)}$ preserves coverage of all previously covered subregions.

\bibliographystyle{IEEEtran}

\bibliography{cas_refs}

\end{document}